\documentclass[aps,prc,twocolumn,showpacs, amsmath,superscriptaddress,floatfix,nofootinbib]{revtex4-1}

\usepackage{setspace,ulem, epsfig,amssymb,amsfonts,amsmath,mathtools,bm,color,xcolor,graphicx,braket,adjustbox,esint,upgreek,verbatim}
\usepackage{ctable}

\newcommand{\bc}[1]{[{\bf \color{blue}{comment}}]}

\begin{document}
%\begin{spacing}{1.0}
%\title{Strong heavy-quark potential from bottomonium sequential suppression in heavy-ion collisions}
\title{
Bottomonium sequential suppression and strong heavy-quark potential in heavy-ion collisions}

\author{Liuyuan Wen} 
\affiliation{Department of Physics, Tianjin University, Tianjin 300350, China}
\author{Baoyi Chen}
\email{baoyi.chen@tju.edu.cn}
\affiliation{Department of Physics, Tianjin University, Tianjin 300350, China}
\date{\today}

\begin{abstract}
We employ the time-dependent Schr\"odinger equation with different complex potentials to study the bottomonium sequential suppression  in Pb-Pb collisions at $\sqrt{s_{NN}}=2.76$ TeV and 5.02 TeV and Au-Au collisions at $\sqrt{s_{NN}}=200$ GeV.  
Both color screening effect and the random scatterings with thermal partons are considered in the real and imaginary parts of the heavy-quark potentials. As the real part of the heavy-quark potential is between the free energy $F(T,r)$ and the internal energy $U(T,r)$ of heavy quarkonium, we parametrize different potentials with a function of $F$ and $U$ to evolve the bottomonium wave packages in the medium. We find that when the real part of the potential is close to $U(T,r)$, it can explain well the pattern of bottomonium sequential suppression where their nuclear modification factors satisfy the relation $R_{AA}(1s)>R_{AA}(2s)>R_{AA}(3s)$ observed in  experiments. In the other limit of  
$F(T,r)$, bottomonium wave packages tend to expand due to weak attracted force, which results in evident transitions from $\Upsilon(2s)$ to $\Upsilon(3s)$ components and does not satisfy the sequential suppression pattern. We suggest that the bottomonium sequential suppression can be a probe of strong  heavy-quark potential in the medium.

\end{abstract}
%\pacs{ 14.40.Nd, 12.38.Mh, 25.75.-q }
\maketitle

\section{Introduction}
Relativistic heavy-ion collisions can generate an extremely hot deconfined medium consisting of quarks and gluons, called Quark-Gluon Plasma (QGP)~\cite{Bazavov:2011nk}. 
Light hadrons are melt into partons with deconfinement phase transitions. While heavy quarkonium with large binding energy may survive in the medium up to a dissociation  temperature $T_d$~\cite{Matsui:1986dk,Satz:2005hx,Zhao:2020jqu}. 
The production and collective flows of heavy quarkonium are regarded as probes of the 
hot medium profile in heavy-ion 
collisions~\cite{Chen:2019qzx,Zhao:2021voa,Liu:2010ej,Strickland:2011mw,Du:2018wsj, Strickland:2011aa,Brambilla:2021wkt,Yao:2021lus}. 
At temperatures below $T_d$, quarkonium 
bound states also suffer 
random scatterings with thermal partons. 
In transport models~\cite{Yan:2006ve,Chen:2016dke,
Zhao:2007hh, Du:2015wha, Yao:2020xzw, Yao:2020eqy,Zhao:2022ggw}, 
color screening 
effect and the inelastic scatterings are considered in the 
collision cross section with thermal partons. While in potential models, 
the effect of random scatterings is approximated to be an imaginary potential on the 
quarkonium wave function~\cite{Blaizot:2018oev,Blaizot:2021xqa,
Katz:2015qja,Gossiaux:2016htk}. 
Both color screening effect and particle scatterings are included in the 
complex potentials of quarkonium. 
The inverse transition from color-octet to color-singlet state is ignored herein,   
which turns out to be
less important in the nuclear 
modification factors of bottomonium~\cite{Akamatsu:2014qsa}. 
There have been some potential models such as the Schr\"odinger-Langevin equation 
which describes the evolution of quarkonium wave functions with the 
stochastic potential directly~\cite{Katz:2015qja}. 
Open quantum system approaches 
are also developed to evolve the density matrix of the subsystems with  
the Lindblad equation formalism~\cite{Brambilla:2016wgg,Brambilla:2020qwo}
 and the stochastic wave packages with the Stochastic Schr\"odinger 
equation~\cite{Akamatsu:2018xim,Xie:2022tzs}.

The sequential suppression pattern of bottomonium states 
have been observed in Pb-Pb collisions at the Large Hadron Collider (LHC)~\cite{CMS:2018zza,CMS:2016rpc,ALICE:2014wnc} and the Relativistic Heavy-ion Collider (RHIC)~\cite{PHENIX:2014tbe,STAR:2013kwk}.  
The dissociation of quarkonium is closely connected with the in-medium heavy quark potentials. 
One can use the sequential suppression pattern to extract the in-medium potential. 
In this work, we take different complex potentials to evolve the 
bottomonium wave 
packages in the hot medium. 
The inner motion of wave packages is described with 
the time-dependent Schr\"odinger equation. 
As the real part of the heavy quark potential is between two limits: the free energy $F$ and the internal energy $U$, a parametrized potential consisting of $F$ and $U$ are taken in the 
Hamiltonian. 
We studied the suppression of bottomonium  states $\Upsilon(1s,2s,3s)$
at LHC energy (2.76 TeV, 5.02 TeV) 
Pb-Pb collisions and the RHIC energy (200 GeV) Au-Au collisions. 
The regeneration process from uncorrelated bottom and anti-bottom quarks in QGP are neglected in the present framework~\cite{Chen:2017duy,Blaizot:2015hya}. 
It turns out to be less important compared with the primordial production.

Firstly, we will introduce the framework of the Schr\"odinger equation model. Then we parametrize the 
complex potentials based on the data from Lattice QCD calculations. The realistic spatial distribution of bottomonium produced in Pb-Pb collisions is proportional to the number of binary collisions. While the initial momentum distribution is extracted from the pp collisions. After bottomonium wave packages move out of the hot medium, the final production of bottomonium states is obtained by projecting the wave packages to the $\Upsilon(1s,2s,3s)$ states which are defined as the eigenstates of the vacuum Cornell potential. The inclusive nuclear modification factors of each state are given after considering the feed-down process from higher to lower states.

\section{Theoretical model}
In the bottomonium evolution, 
hot medium effects are included in the temperature-dependent complex potential. 
Neglect the viscosity of the medium, 
both real and imaginary potentials become central. There is no mixing between the 
states with different angular momentum in the 
evolution of bottomonium wave package. 
Due to the large mass of bottom quarks compared with the inner motion of bottomonium, one can neglect the relativistic effect. 
We separate the radial part of the Schr\"odinger equation for bottomonium evolution~\cite{Wen:2022utn}, 
\begin{align}
\label{fun-rad-sch}
i\hbar {\partial \over \partial t}\psi( r, t) = [-{\hbar^2\over 2m_\mu}{\partial ^2\over \partial r^2} +V( r, T) + {L(L+1)\hbar^2\over 2 m_\mu r^2}]\psi(r,t)
\end{align}
where $r$ is the radius of the wave package. 
$t$ is the proper time 
in the local rest frame of bottomonium.  
The reduced mass is defined as 
$m_\mu=m_1m_2/(m_1+m_2)=m_b/2$ with bottom mass to be $m_b=4.62$ GeV. The wave package  
$\psi(r,t)=rR(r,t)$ is a product of the $r$ and the radial wave function $R(r,t)$. 
It is regarded as a quantum superposition of 
bottomonium eigenstates of the vacuum Cornell potential, ${\psi(r,t)\over r}=\sum_{nl} c_{nl}(t) \Phi_{nl}(r)$. $\Phi_{nl}(r)$ is the radial wave 
function of the eigenstate with the fixed radial 
and angular quantum number $(n,l)$. The 
square of the coefficient $|c_{nl}(t)|^2$ is 
the fraction of the corresponding eigenstate in the 
wave package. It changes with time due to 
the complex heavy-quark potential $V(r,T)$. 

In the hot medium, the real part of the heavy quark 
potential is reduced by the color screening. 
It is between the free energy $F(T,r)$ and the internal energy $U(r,T)$. 
Some studies have indicated a strong attracted potential of 
quarkonium in the medium~\cite{Wen:2022utn,Liu:2010ej, Du:2019tjf}. 
We parametrize 
the free energy with the form~\cite{Islam:2020bnp},
\begin{align}
    F(r,T)=-{\alpha\over r}e^{-m_Dr} +{\sigma\over m_D}(1-e^{-m_D r})
\end{align}
where $m_D=T\sqrt{{4\pi N_c\over 3}\alpha (1+{N_f\over 6})}$ 
is the in-medium gluon Debye 
mass~\cite{Islam:2020bnp}. 
The values of $\alpha$ and $\sigma$ are determined in the Cornell potential with the masses of bottomonium states (1S,1P), $\alpha=\pi/12$, $\sigma=0.2\ \rm{(GeV)}^2$, {where 
the corresponding vacuum masses of the states are $m_{1S,1P}=(9.47,9.79)$ GeV~\cite{Satz:2005hx}. } The factors of color and flavor are taken as $N_c=N_f=3$. The internal energy is obtained via the relation 
$U(r,T)=F+T(-\partial F/\partial T)$,
\begin{align}
U(r,T) &= -{\alpha\over r}(1+m_D r)e^{-m_D r} \nonumber \\
&\quad + {2\sigma \over m_D}[1-e^{-m_D r}] - \sigma re^{-m_D r}
\end{align}
In the following calculations, we parametrize 
the real potential with a combination of $F$ and $U$, $V_R=xF+(1-x)U$. The parameter $x\in [0,1]$ indicates 
that the real potential is between the free energy and the internal energy. 
The dependence of $(r,T)$ in the parameter $x$ has been 
temporarily neglected. 

\begin{figure}[tbp]
\centering 
\includegraphics[width=0.35\textwidth]{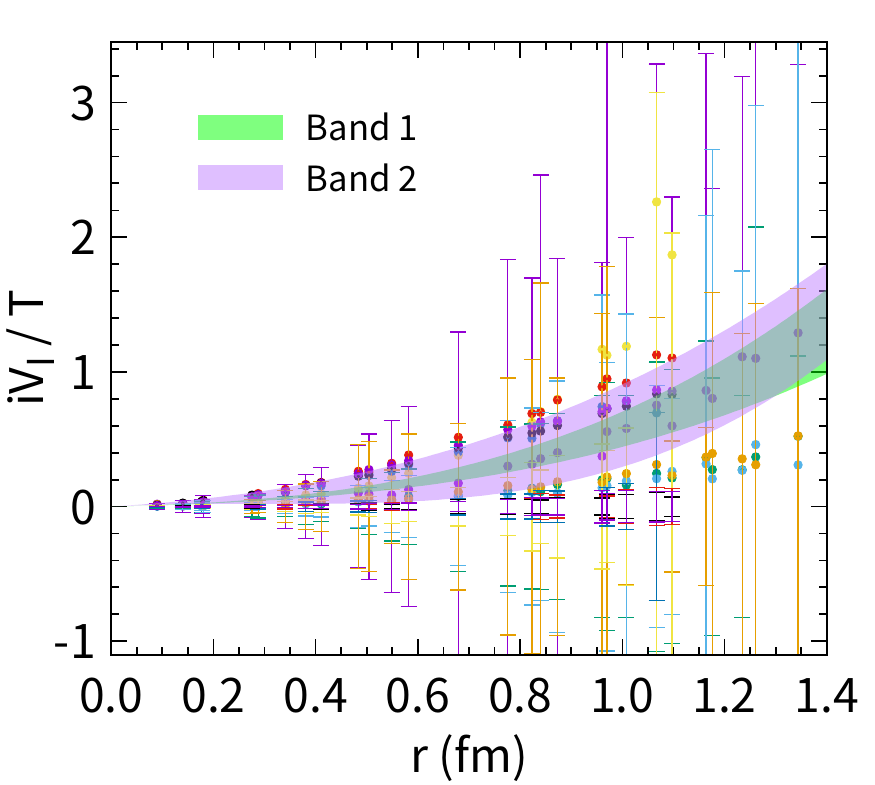}
\caption{\label{fig:ImagV} 
The imaginary part of the heavy quark potential scaled with the temperature $iV_I/T$ as a function of the radius. Different uncertainties of the data points are partially considered with ``Band 1'' and ``Band 2''. The potential data is cited from~\cite{Burnier:2016mxc}.
}
\end{figure}

Random scatterings from thermal partons contribute noise 
terms in the potential. The stochastic evolution of the 
wave package with a noisy potential 
is well-approximated by the evolution of the averaged 
wave function with complex potential~\cite{Wen:2022utn,Islam:2020gdv}. 
According to the results from lattice QCD calculations, 
the imaginary potential shows near-linear 
dependence on the temperature. 
Therefore, we fit the lattice 
data points with the polynomial up to cubic terms of the radius~\cite{Shi:2021qri}. Due to the large uncertainty in the data points, we perform two kinds of $V_I$ parametrization for the following calculations. Firstly, we fit the central values of the data points at $r<1.2$ fm, while data points at a very large radius have been neglected due to extremely large uncertainty. This line is then shifted upward 
to partially include the uncertainty in the data points~\cite{Wen:2022utn}. 
Two lines are plotted as the lower and upper 
limits of the ``Band 1'' in Fig.\ref{fig:ImagV}, where the expressions of two lines are,
\begin{align}
\label{eq-VI}
   & V_I^{\rm upper}\left(\bar{r},T\right)=-iT(a_1\bar{r}^3+a_2\bar{r}^2+a_3\bar{r})\nonumber \\
   & V_I^{\rm lower}\left(\bar{r},T\right)=-iT(b_1\bar{r}^3+b_2\bar{r}^2+b_3\bar{r}),
\end{align}
where $a_1=0.2096,a_2=0.1486, a_3=0.0854$ and 
$b_1=0.3605, b_2=0.2536, b_3=0.0909$. $i$ is an imaginary number. ${\bar r}\equiv r/fm$ is the dimensionless variable. 
%The upper and lower limits of ``Band 1''
%$V_I^{\rm upper}$ and $V_I^{\rm lower}$ consider some of the uncertainty in the data points in Fig.\ref{fig:ImagV}.  
In order to consider a more uncertainty in the data points, we use all the 
data points to fit a line and then consider one standard deviation error bar in the fit. This will give a larger band labeled with ``Band 2'' in Fig.\ref{fig:ImagV}. The larger uncertainty in the $V_I$ will result in significant uncertainty in bottomonium $R_{AA}$. In the following calculations about bottomonium, $V_I$ labeled with ``Band 1'' is firstly employed. In the last section, $V_I$ labeled with ``Band 2'' is also considered to study the uncertainty of model parameters.

The time-dependent Schr\"odinger equation with complex potentials can be solved numerically with the Crank-Nicolson method, where the wave package is directly evolved in the coordinates instead of projecting to a series of eigenstates. The discrete form of the radial Schr\"odinger equation becomes (with natural unit $\hbar=c=1$), 
\begin{align}
  \label{eq-sch-num}
{\bf T}_{j,k}^{n+1}\psi_{k}^{n+1} = \mathcal{V}_{j}^{n}.
\end{align}
Here $j$ and $k$ are the index of rows and columns in the matrix $\bf T$ respectively. 
The non-zero elements in the matrix are, 
\begin{align}
\label{eq-sim-cn}
&{\bf T}^{n+1}_{j,j}= 2+2a+bV_j^{n+1}, \nonumber \\
&{\bf T}^{n+1}_{j,j+1}={\bf T}^{n+1}_{j+1,j}= -a, \nonumber \\
&\mathcal{V}_j^n= a\psi_{j-1}^n +(2-2a-bV_j^n)\psi_j^n +a\psi_{j+1}^n ,
\end{align}
the parameters are $a= i\Delta t/(2m_\mu (\Delta r)^2)$, and $b=i\Delta t$. $\Delta t=0.001$ fm/c and $\Delta r=0.03$ fm is the step of the time and the radius. 
The subscript $j$ and the superscript $n$ represents the 
coordinate and the time, $r_j=j\cdot \Delta r$ and 
$t^n=n\cdot \Delta t$ respectively. As the 
potential depends on the temperature of the medium which 
changes with the time and the coordinate, hamiltonian 
depends on time. 
At each time step, we calculate the inverse of the matrix $\bf{T}$ to get the wave package at the next time step. 
As the medium local temperature only changes evidently after a time scale larger than $\sim 0.1$ fm/c, 
We can approximately use the same value of the matrix in this time scale, and update the value of the 
matrix and its inverse 
after a time period of $100\Delta t$. This approximation significantly reduces the 
numerical computation time. 
The fractions of the bottomonium eigenstates in the wave package of the bottom dipole is obtained by projecting 
$\psi(r,t)$ to the wave functions of the corresponding 
eigenstates $\Phi_{nl}$. 

In Pb-Pb collisions, bottom pairs are produced in parton hard 
scatterings, 
{where the initial wave functions of bottom pairs 
are close to a Delta function in coordinates at the time scale $\tau\sim 1/(2m_b)$. As 
bulk medium are described with the hydrodynamic equations from 
$\tau_0 \sim 0.6$ fm/c~\cite{Zhao:2017yhj}, where the medium reaches 
local equilibrium, we start evolving the Schr\"odinger equation from $\tau\ge \tau_0$ and assume that the bottom pairs have 
evolved into bottomonium eigenstates in the time scale $\tau\sim \tau_0$. In the Schr\"odinger equation, the initial conditions of wave packages at 
$\tau=\tau_0$ are taken 
as the bottomonium vacuum eigenstates.  }
 The mixing between different eigenstates in the 
 wave package will be induced by the in-medium potential which corresponds to the 
transitions between different eigenstates. 
With multiple wave packages generated at different positions, they move along different trajectories in the hot medium and experience different temperature profiles. We randomly generate a large set of wave 
packages representing the primordially produced 
bottomonium, and evolve their wave packages event-by-event with the Schr\"odinger equation. The ensemble-averaged final fractions of the bottomonium eigenstates in the 
wave package are obtained and used to calculate the direct and inclusive
nuclear modification factor $R_{AA}(1s,2s,3s)$.

\section{initial conditions and hot medium evolutions}
In proton-proton (pp) collisions, experiments have measured the production cross section of $\Upsilon(1s,2s,3s)$ respectively. The inclusive cross sections $\sigma_{\rm exp}$ 
of $\Upsilon(1s,1p,2s,2p,3s)$ with the 
feed-down process are listed in 
table~\ref{tab-lab}. The direct cross section without the feed-down process is connected with the experimental data with the branching ratios from the 
particle data group~\cite{ParticleDataGroup:2018ovx}. {It satisfies the relation  $\sigma_{\rm exp}(1s)=\sigma_{\rm direct}(1s)+\sum_{nl}\sigma_{\rm direct}(nl)\mathcal{B}_{nl\rightarrow 1s}$}. The radial and angular quantum number $n,l$ represents the states of (1p,2s,2p,3s) when taking different values. In the calculations, We treat the 
$\chi_{b0,b1,b2}(1p)$ to be the same 1p state where the branching ratio of $\mathcal{B}_{1p\rightarrow 1s}$ is the average of the three branching ratios 
 $\mathcal{B}(\chi_{b0,b1,b2}(1p)\rightarrow 1s)$. The potential for the $\chi_{b0,b1,b2}(1p)$ is also the same in the Schr\"odinger equation.
Similarly, 
$\chi_b(2p)$ represents the sum of $\chi_{b0,b1,b2}(2p)$ states. For the inclusive cross sections of excited states, we can obtain a similar relation $\sigma_{\rm exp}(2s)=\sigma_{\rm direct}(2s)+\sum_{nl}\sigma_{\rm direct}(nl)\mathcal{B}_{nl\rightarrow 2s}$ where $n,l$ here represents higher states above $\Upsilon(2s)$. With this method we extract the direct cross sections of bottomonium states listed in table \ref{tab-lab}. 
In the initial conditions of the Schr\"odinger 
equation, the number of wave packages initialized as different bottomonium 
eigenstates at $t=\tau_0$ satisfy the ratio $\sigma_{\rm direct}^{1s}:\sigma_{\rm direct}^{1p}:\sigma_{\rm direct}^{2s}:\sigma_{\rm direct}^{2p}:\sigma_{\rm direct}^{3s}$ in table~\ref{tab-lab}.

\begin{table}[htbp]
\caption{Bottomonium production cross section measured in $\sqrt{s_{NN}}=5.02$ TeV pp collisions. 
$\sigma_{\rm exp}$ and $\sigma_{\rm direct}$ represent the cross sections with and without the 
feed-down process. The values of $\sigma_{\rm exp}$ are cited from~\cite{CMS:2013qur,ATLAS:2012lmu,CMS:2010wld,LHCb:2012aa,LHCb:2014dei,LHCb:2014ngh,ALICE:2015pgg,LHCb:2013itw}. 
}
\label{tab-lab}
    \centering
\begin{tabular}{|c|c|c|c|c|c|c|c|c|c|}
\hline State & $\Upsilon(1s)$ & $\chi_{b}(1p)$ & $\Upsilon(2s)$ & $\chi_{b}(2p)$ & $\Upsilon(3s)$ \\
\hline $\boldsymbol{\sigma}_{\text {exp}}(nb)$ & $57.6$ & $33.51$& $19$& $29.42$ & $6.8$  \\
\hline $\boldsymbol{\sigma}_{\text {direct}}(nb)$ & $37.97$ & $44.2$  & $18.27$& $37.68$ & $8.21$ \\
\hline
\end{tabular}
\end{table}

After initializing the wave function of bottom pair to be the bottomonium eigenstates at $t=\tau_0$, the relative motion of the wave packages are controlled by the 
Schr\"odinger equation. While the 
center of the wave packages move in the medium 
with a constant total 
momentum $p_T$. We assume that the momentum 
distribution of 
wave packages satisfy the bottomonium momentum 
distribution measured in pp collisions. 
The normalized initial 
transverse momentum distribution of $\Upsilon(1s)$ in 5.02 TeV 
pp collisions is parametrized with the form, 
\begin{equation}
\label{eq:pp-input}
{dN_{pp}^{\Upsilon}\over d\phi p_T dp_T} = 
{(n-1)\over \pi (n-2) \langle p_T^2\rangle_{pp}} [1+{p_T^2\over (n-2) \langle p_T^2
\rangle_{pp}}]^{-n} 
\end{equation}
where the parameter is fitted to be $n=2.5$, and the 
mean transverse momentum square at the central rapidity
is estimated to be 
$\langle p_T^2\rangle_{pp}=(80,55,28)\ \mathrm{(GeV/c)^2}$ at 5.02 TeV, 2.76 TeV and 200 GeV respectively 
based on the 
measurements in pp collisions~\cite{CDF:2001fdy,LHCb:2014dei,LHCb:2012aa}.
Spatial distributions of 
the wave package is proportional to the number of binary collisions in the overlap of two nuclei. We assume the same distribution for 
bottomonium excited states as the ground state. 
The cold nuclear matter effects such as the shadowing effect also changes the initial distributions of the wave packages. We take the 
EPS09 NLO model to calculate the shadowing factor $\mathcal{R}_{S}({\bf x}_T)$ 
of bottomonium~\cite{Eskola:2009uj}. We randomly generate the 
wave packages based on the spatial distribution, 
\begin{align}
    {dN^{\Upsilon}_{AA}\over d{\bf x}_T} \propto 
    T_A({\bf x}_T+{\bf b}/2)T_B({\bf x}_T-{\bf b}/2) \mathcal{R}_{S}({\bf x}_T)
\end{align}
where $T_{A(B)}$ is the thickness function. ${\bf b}$ is the impact parameter. The center of the wave packages move in the hot medium with a constant momentum satisfying the distribution 
Eq.(\ref{eq:pp-input}).

%\subsection{Hydrodynamic medium}
For the hot medium evolutions, we use the (2+1) dimensional ideal hydrodynamic model developed 
by Tsinghua Group. The hydrodynamic model have been applied in the transport model 
to explain 
well both charmonium and bottomonium observables at RHIC~\cite{Liu:2010ej,Liu:2009wza} and LHC~\cite{Chen:2013wmr,Chen:2018kfo,Shi:2017qep} energies. 
At $\sqrt{s_{NN}}=5.02$ TeV Pb-Pb collisions, 
the maximum initial temperature at the 
center of the fireball is determined to be $T_0({\bf x}_T=0)=510$ MeV at the time 
$\tau_0=0.6$ fm in central rapidity. The spatial distributions of the initial energy density 
is obtained with the optical Glauber initial condition. The equation of state of the QGP and 
the hadron gas are taken as the ideal gas and hadron resonance gas, respectively. 
The phase transition between two phases is a first-order transition with the critical 
temperature determined to be $T_c=165$ MeV by the bag model. Wave packages start evolution with the parametrized
complex potentials from $\tau\ge \tau_0$ along different trajectories. When the local 
temperature along quantum trajectories 
is smaller than $T_c$, 
in-medium potential is replaced with the 
Cornell potential, where the fraction of the bottomonium eigenstates in wave package 
do not change any more. With this approximation, 
we have neglected the hadron gas effects, which is believed to be small 
for tightly bound bottomonium. 

\section{Observables} %RAA

At each time step, the dynamical evolution of the wave package is described with the Schr\"odinger equation where the hot medium effects and the cold nuclear matter effects are encoded in the complex potential and the initial conditions respectively. The wave packages move with a constant velocity in the medium. The final fractions of the bottomonium eigenstates in the wave packages are calculated when the center of wave packages move out of the QGP. After ensemble average over a large set of wave packages generated with different initial positions and the total momenta, we get the mean fraction $|c_{nl}(t)|^2$ of the bottomonium eigenstate labelled with the quantum number $(n,l)$ in the wave package to be,

\begin{align}
\label{eq-dRAA}
\langle |c_{nl}(t)|^2\rangle_{\rm en} = 
{\int d{\bf x}_{\Upsilon}d{\bf p}_{\Upsilon} |c_{nl}(t, {\bf x}_{\Upsilon}, 
{\bf p}_{\Upsilon})|^2{{dN^{\Upsilon}_{AA}}\over d{\bf x}_{\Upsilon} d{\bf p}_{\Upsilon}}
\over 
\int d{\bf x}_{\Upsilon}d{\bf p}_{\Upsilon} 
{ {dN^{\Upsilon}_{AA}}\over d{\bf x}_{\Upsilon} d{\bf p}_{\Upsilon}}}
\end{align}
where ${\bf x}_{\Upsilon}$ and ${\bf p}_{\Upsilon}$ are the position and the total 
momentum of the center of a wave package respectively.  
$dN_\Upsilon/d{\bf x}_\Upsilon d{\bf p}_\Upsilon$ represents the distribution 
of wave packages in phase space. After the evolution of the Schr\"odinger equation, excited states decay into 
the ground state and contribute to the inclusive $R_{AA}$ of $\Upsilon(1s)$ and $\Upsilon(2s)$,  
\begin{align}
\label{eq-promptRAA}
R_{AA}(1s)
= {\sum_{nl} \langle |c_{nl}(t)|^2\rangle_{\rm en} f_{pp}^{nl}
\mathcal{B}_{nl\rightarrow 1s}\over \sum_{nl}
\langle |c_{nl}(t_0)|\rangle^2\rangle_{\rm en}  f_{pp}^{nl} \mathcal{B}_{nl\rightarrow 1s}}
\end{align}
$\mathcal{B}_{nl\rightarrow 1s}$ is the branching ratio of the eigenstate ($n,l$) decaying into the 1s state. The values of branching ratios are taken from the particle data group. 
In the calculation of inclusive $R_{AA}(1s)$, all the decay channels of 
$\chi_b(1p)\rightarrow \Upsilon(1s),\Upsilon(2s)\rightarrow \Upsilon(1s),\chi_b(2p)\rightarrow \Upsilon(1s),\Upsilon(3s)\rightarrow \Upsilon(1s)$ are included in Eq.(\ref{eq-promptRAA}). For inclusive $R_{AA}(2s)$, 
the decay contribution from $\chi_b(2p),\Upsilon(3s)$ are considered. There is no higher states 
decaying into $\Upsilon(3s)$.

\section{Numerical results}

In heavy-ion collisions, we take different complex potentials in the Schr\"odinger equation to calculate the nuclear modification factors of bottomonium at LHC and RHIC energies. 
The real part of the potential is parametrized with a function of $F$ and $U$. While the imaginary potential 
is parametrized with a polynomial function. 
Different uncertainties in the imaginary potential are 
considered with the ``Band 1'' and ``Band 2'' in Fig.\ref{fig:ImagV}. In this section, $V_I$ with 
a smaller uncertainty is taken into the calculations.

%\begin{comment}
%\begin{widetext}
\begin{figure}[tbp]
\centering 
\includegraphics[width=0.23\textwidth]{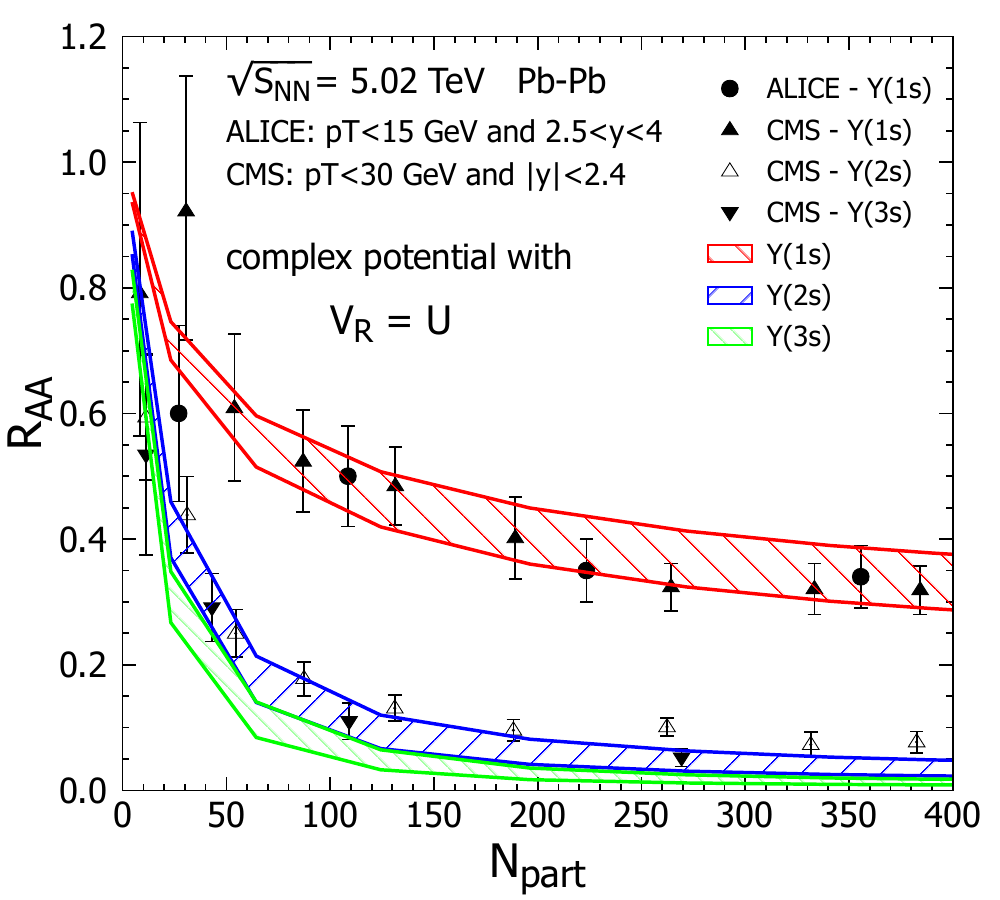}
%\hbox{ } \hspace{0.3cm} \hbox{ }
\includegraphics[width=0.23\textwidth]{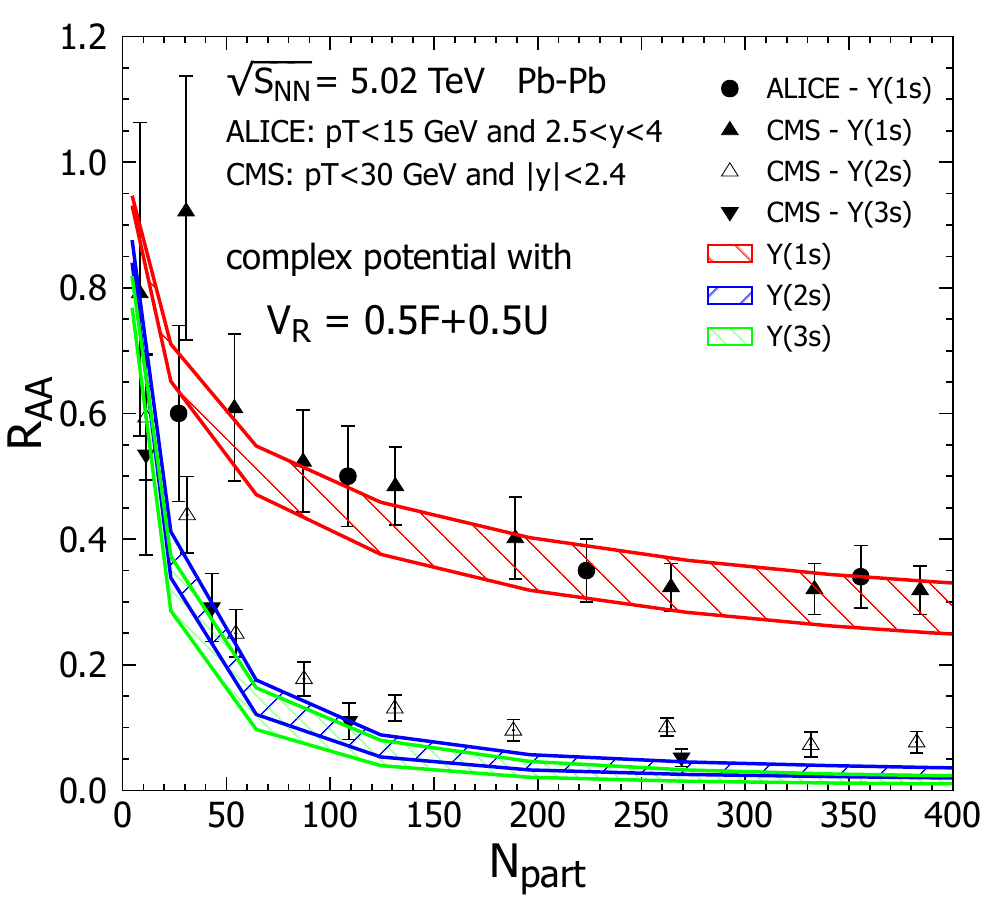}
%\hbox{ } \hspace{0.3cm} \hbox{ }
\includegraphics[width=0.23\textwidth]{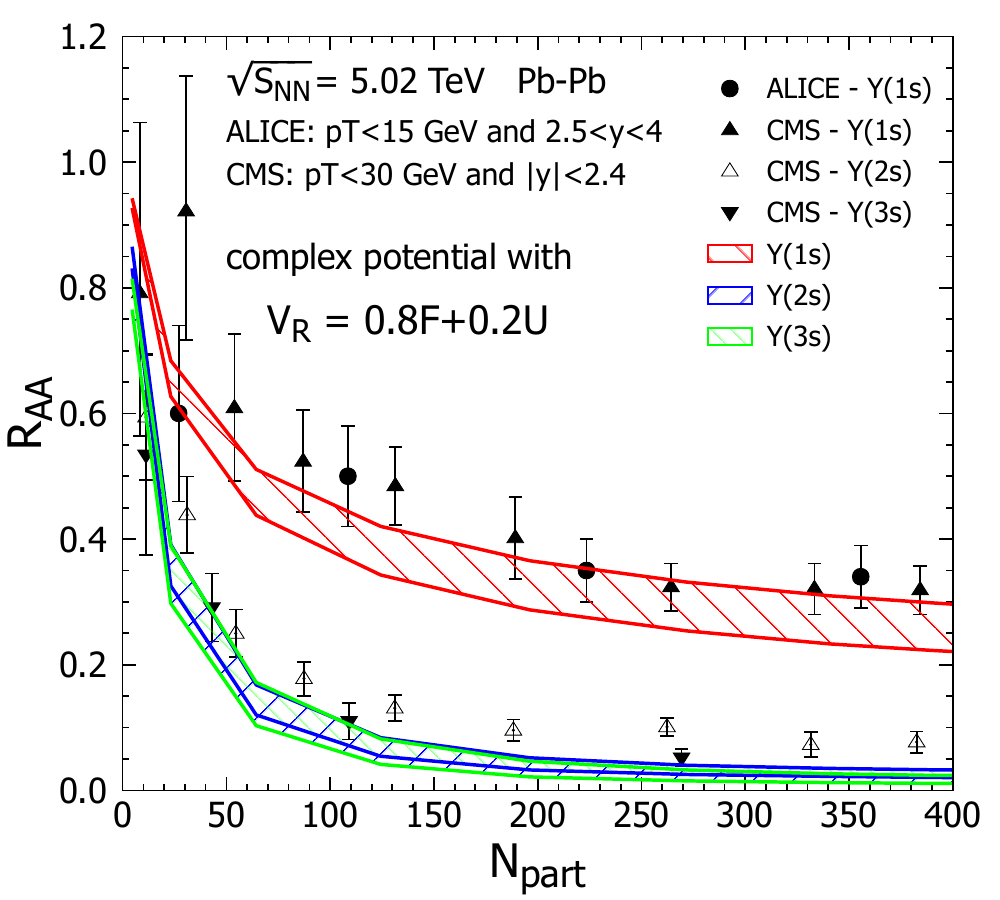}
%\hbox{ } \hspace{0.3cm} \hbox{ }
\includegraphics[width=0.23\textwidth]{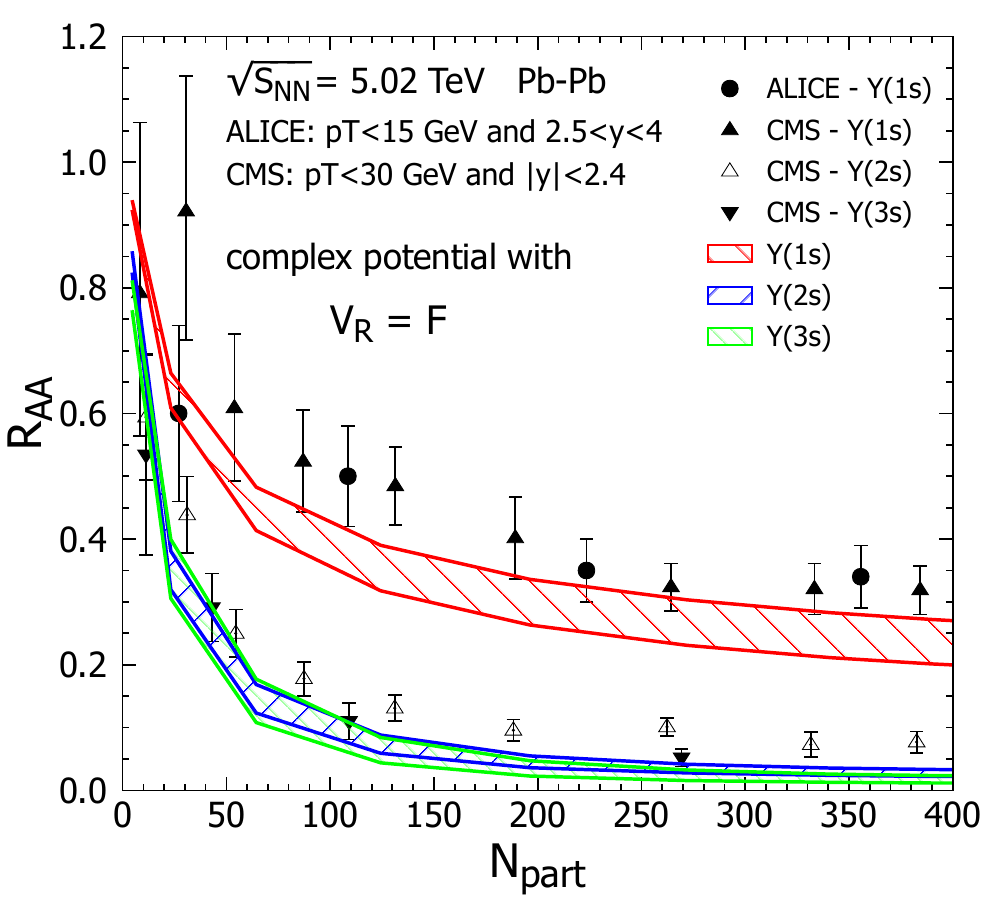}
\caption{\label{fig:RAA-Np-FU} 
The nuclear modification factors of bottomonium 
states $\Upsilon(1s,2s,3s)$ as a function of the number of participant $N_p$ in $\sqrt{s_{NN}}=5.02$ TeV Pb-Pb collisions. 
The complex potentials are taken where the real part of the potential is parametrized with the form $V_R=xF+(1-x)U$, with $x=(0, 0.5,0.8,1)$ respectively in above panels.  The band of the theoretical results corresponds to the uncertainty of the imaginary potential fitted before. 
The experimental data is cited from ALICE~\cite{ALICE:2018wzm} and CMS~\cite{CMS:2017ycw} Collaborations. 
}
\end{figure}
%\end{widetext}
%\end{comment}

In Fig.\ref{fig:RAA-Np-FU}, the suppression of bottomonium $R_{AA}$ as a function of the number of participant $N_p$ are mainly induced by the imaginary potential and the color screening effect. In the case of $V_R=U$, strong heavy quark potential can constrain the wave package. The theoretical calculations show a clear pattern of the sequential suppression, $R_{AA}^{1s}>R_{AA}^{2s}>R_{AA}^{3s}$ in Fig.\ref{fig:RAA-Np-FU}. The bands 
of the theoretical results come from the uncertainty in the imaginary potential ($V_I^{\rm upper}$ and $V_I^{\rm lower}$). For excited states such as $\Upsilon(3s)$, its geometry size is larger than the ground state. As $V_I$ increases with the distance and the 
temperature, bottomonium excited states suffer 
stronger dissociation than the ground state. 
However, when the heavy quark potential is close to the free energy, such as in the cases of $V_R=F$ and $V_R=0.2U+0.8F$,  the attracted force in the wave package becomes weak. The wave package tends to expand outside. The components of the ground state tend to transit into the excited states. In the weak heavy quark potential with $V_R=F$ in Fig.\ref{fig:RAA-Np-FU}, the theoretical band of $R_{AA}^{1s}$ is smaller than the experimental data. Besides, 
the pattern of of $\Upsilon(2s,3s)$ sequential suppression is not evident anymore.

%\begin{widetext}
\begin{figure}[tbp]
\centering 
\includegraphics[width=0.23\textwidth]{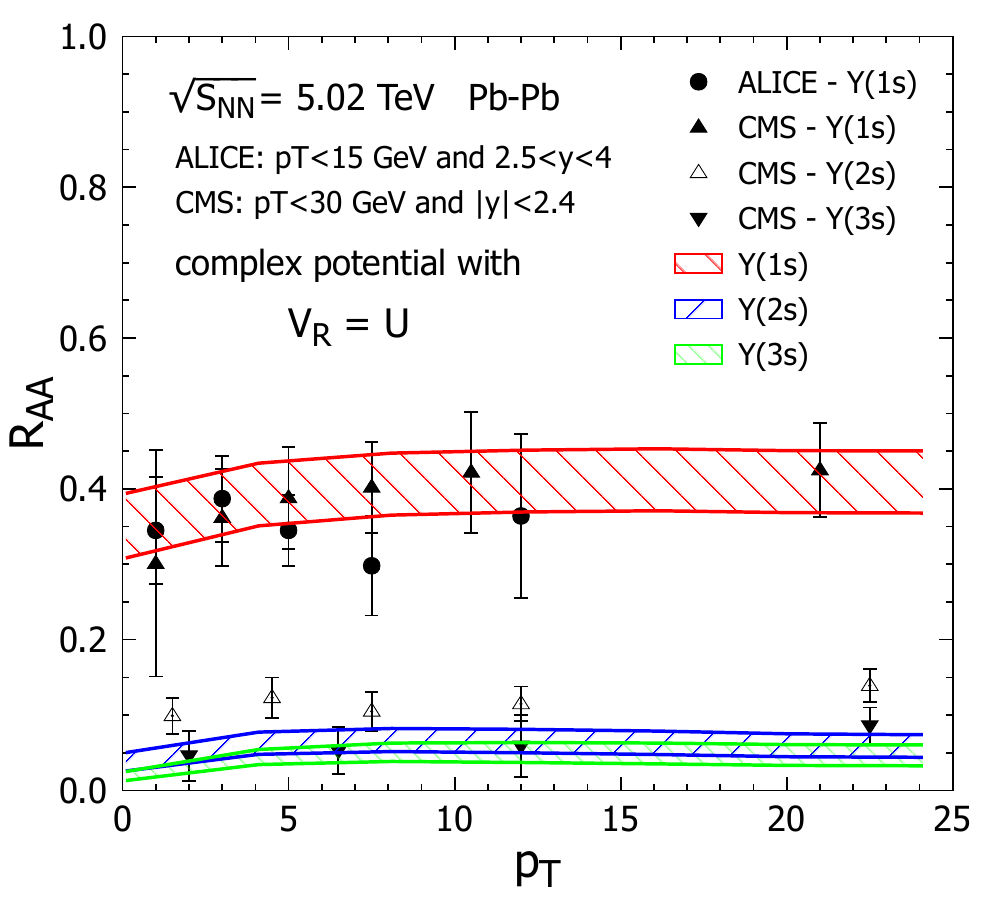}
\includegraphics[width=0.23\textwidth]{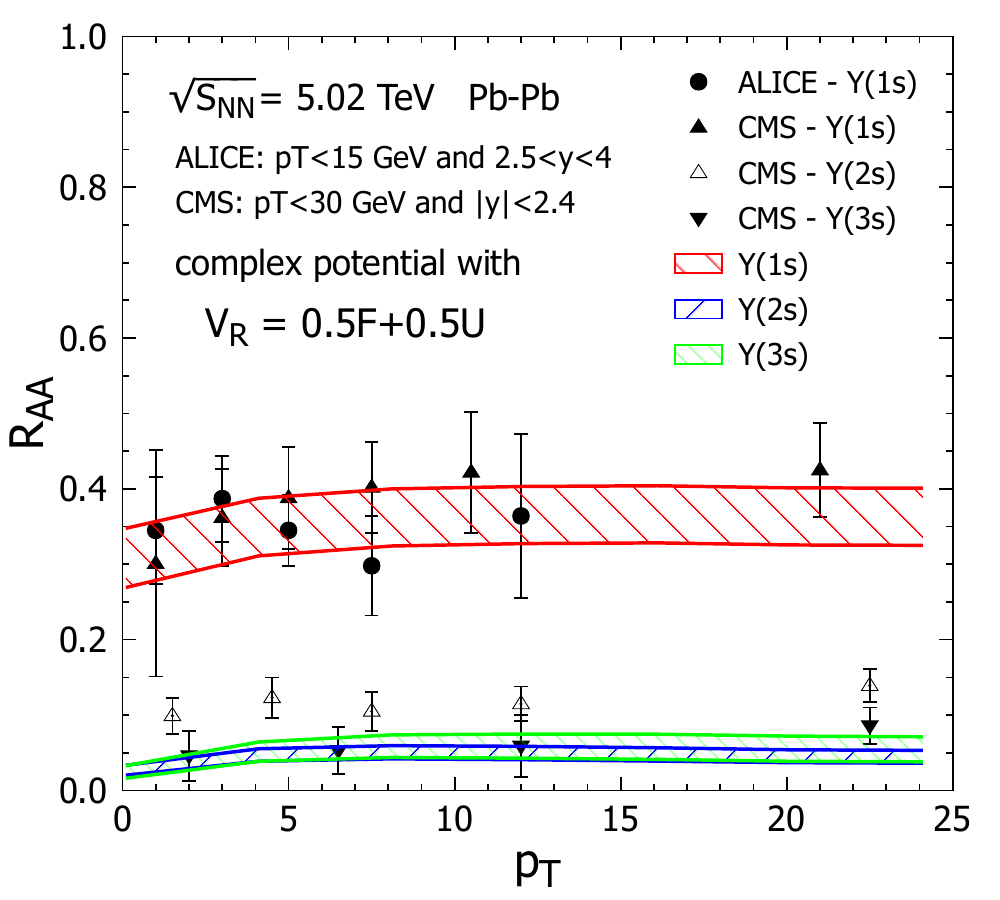}
\includegraphics[width=0.23\textwidth]{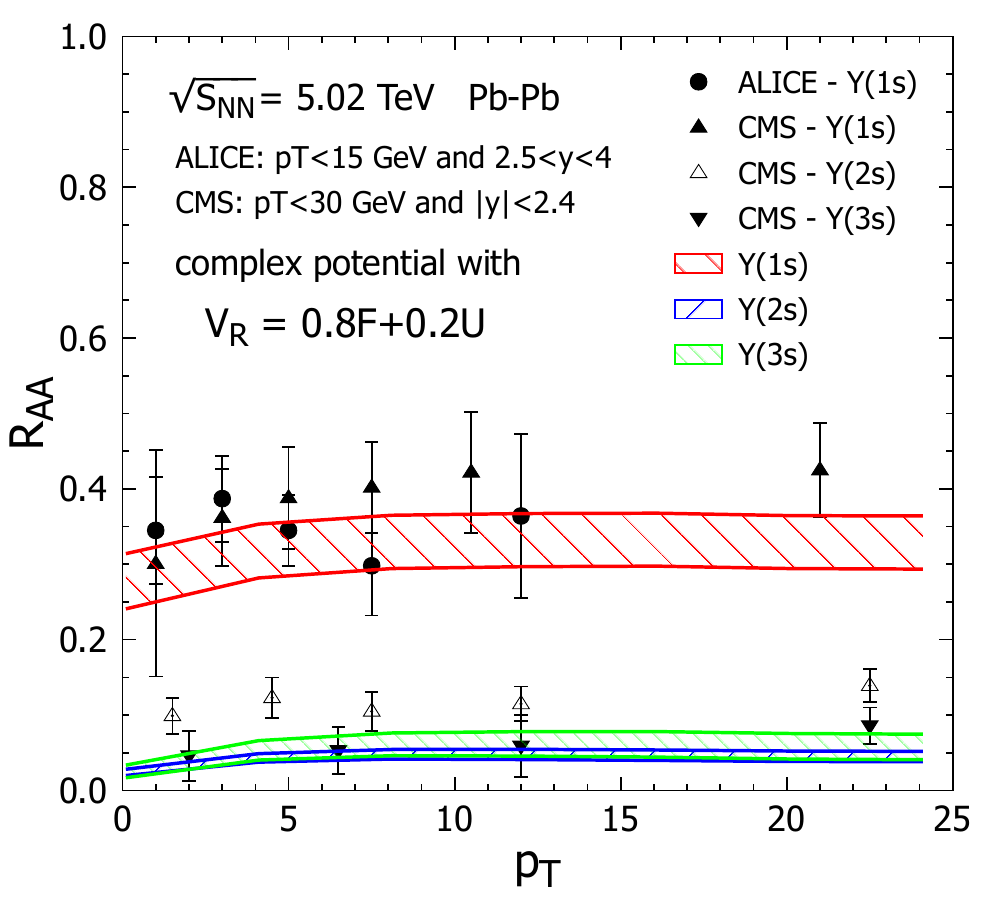}
\includegraphics[width=0.23\textwidth]{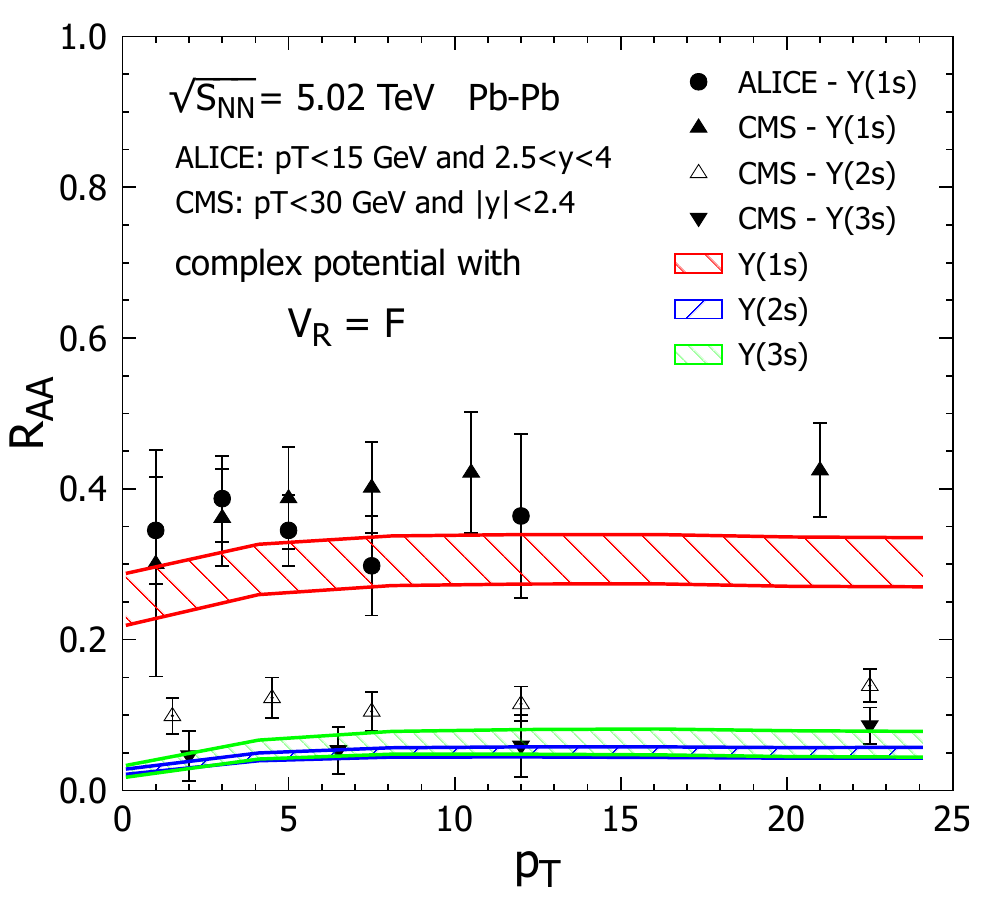}
\caption{\label{fig:RAA-pT-FU} 
The nuclear modification factors of bottomonium 
states $\Upsilon(1s,2s,3s)$ as a function of $p_T$ in minimum-bias $\sqrt{s_{NN}}=5.02$ TeV Pb-Pb collisions. 
The complex potentials are taken, where the real part of the potential is parametrized with the form $V_R=xF+(1-x)U$, with $x=(0,0.5,0.8,1)$ respectively in above panels.  The band of the theoretical results corresponds to the uncertainty of the imaginary potential fitted before. 
The experimental data is cited from ALICE~\cite{ALICE:2018wzm} and CMS~\cite{CMS:2017ycw} 
Collaborations.  
}
\end{figure}
%\end{widetext}

For the $p_T$-differential $R_{AA}$ in Fig.\ref{fig:RAA-pT-FU}, there is a clear sequential suppression pattern in $R_{AA}(1s,2s,3s)$ with $V_R=U$. The theoretical bands can qualitatively explain the experimental data of $\Upsilon(1s,2s,3s)$. The slight increase of $R_{AA}(p_T)$ with $p_T$ is because of the fact that wave packages with larger momentum tend to escape from the hot medium with a shorter time, where the suppression is weaker. 
While with a 
weak heavy quark potential ($V_R=F$), the quantum transition in the wave package makes $R_{AA}^{3s}$ become larger than the value of $\Upsilon(2s)$. This is due to the expansion of 
wave package, which increases the overlap between the wave package and the $\Upsilon(3s)$ wave function. The uncertainty in the imaginary potential is partially included in $R_{AA}$ with a band. 
It does not change the conclusion about the sequential suppression and the strong heavy quark potential. 
Therefore, the sequential suppression pattern of bottomonium indicates a strong heavy quark potential in the hot medium. 
The exact $r-$ and $T-$dependence in the parameter $x$ in the parametrization of $V_R$ is beyond the scope of present work and will be left in following works.

\begin{figure}[tbp]
\centering 
\includegraphics[width=0.23\textwidth]{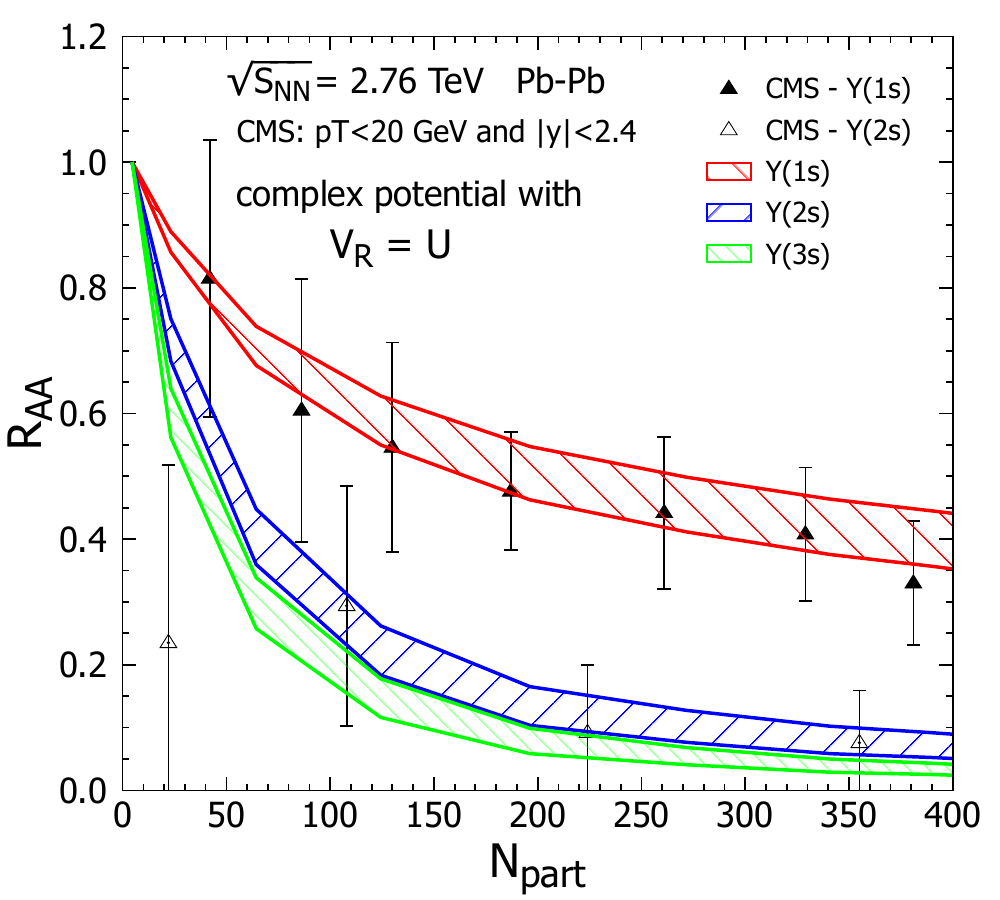}
\includegraphics[width=0.23\textwidth]{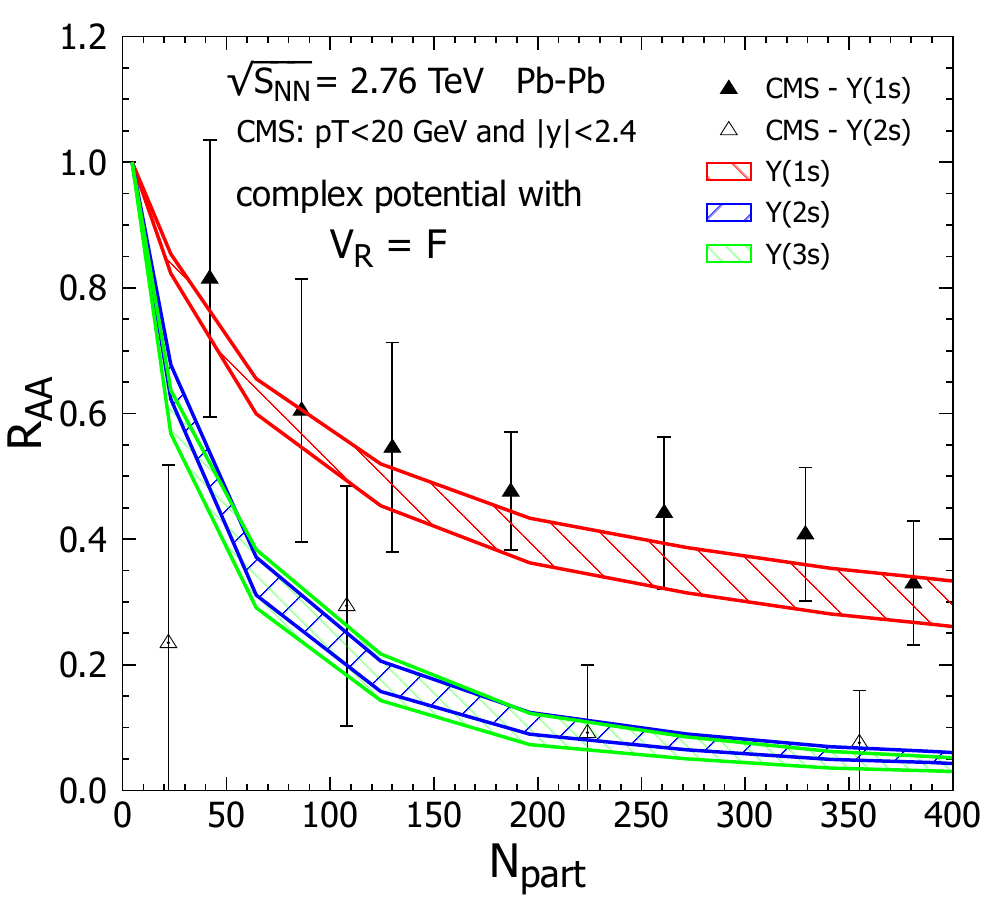}
\includegraphics[width=0.23\textwidth]{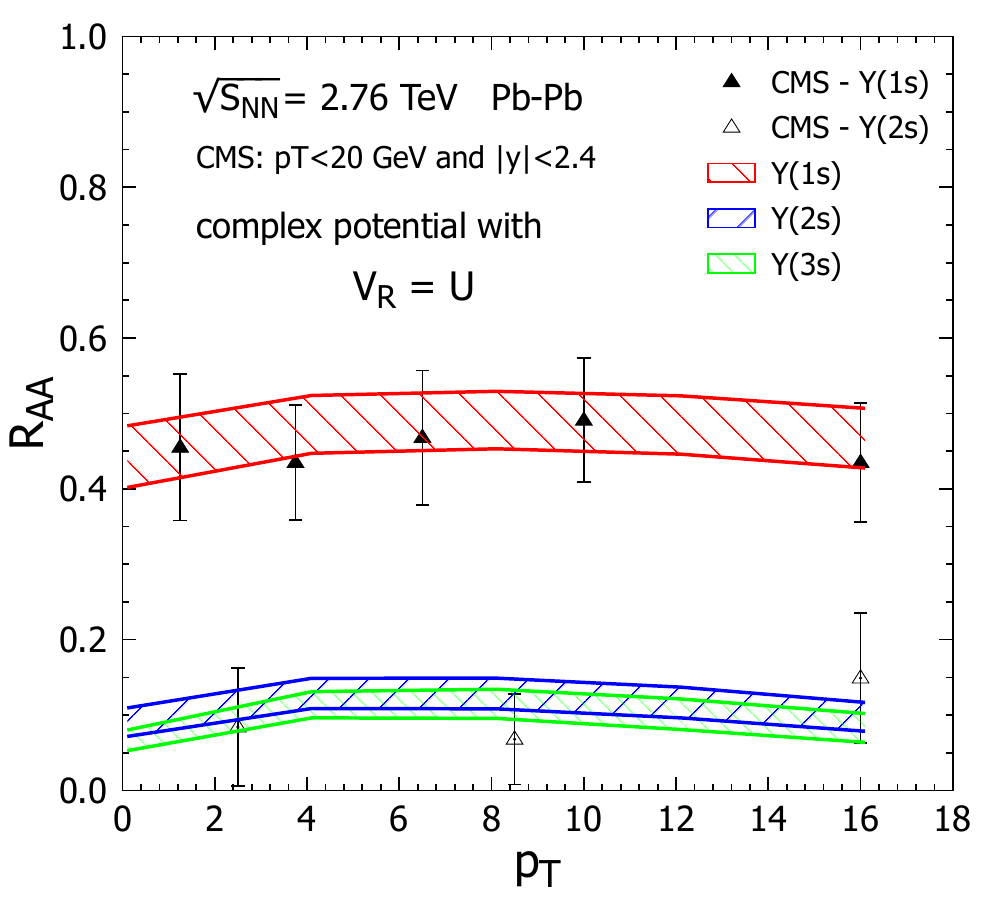}
\includegraphics[width=0.23\textwidth]{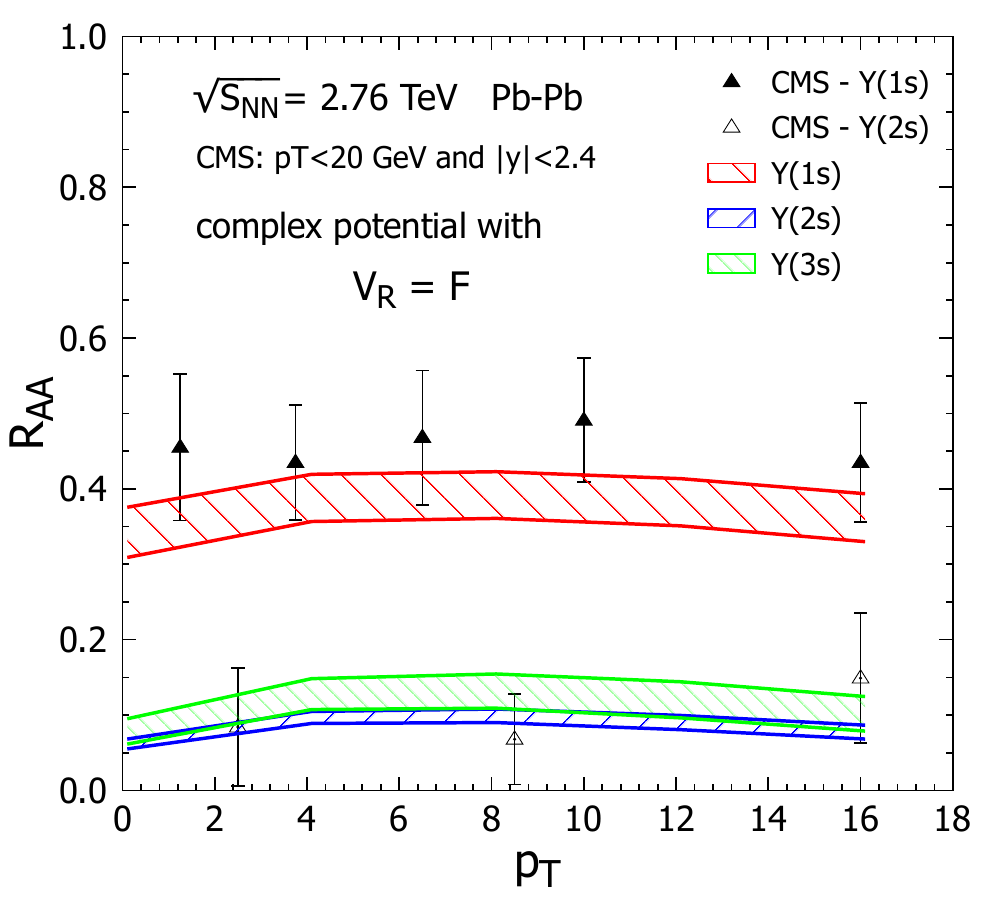}
\caption{\label{fig-2760RAA-Np} 
The nuclear modification factors of bottomonium 
states $\Upsilon(1s,2s,3s)$ as a function of $N_p$ and $p_T$ in $\sqrt{s_{NN}}=2.76$ TeV Pb-Pb collisions. $R_{AA}(p_T)$ is in minimum-bias. 
The real part of the potential is taken as $V_R=U$ and $V_R=F$ respectively.  The band of the theoretical results corresponds to the uncertainty of the imaginary potential. 
The experimental data is cited from CMS
Collaboration~\cite{CMS:2016rpc}.  
}
\end{figure}

Next we turn to the experimental data about 
bottomonium at $\sqrt{s_{NN}}=2.76$ TeV Pb-Pb collisions. The experimental data in Fig.\ref{fig-2760RAA-Np} is cited from CMS Collaboration. At $\sqrt{s_{NN}}=2.76$ TeV, the production cross section of $\Upsilon(1s,2s)$ are extracted to be $d\sigma_{\Upsilon}/dy=(30.3,10)$ nb respectively based on the measurement in the central rapidity pp collisions~\cite{Du:2017qkv,CMS:2016rpc}. While the production cross section of $\Upsilon(3s)$ is extracted with the same ratio of $\sigma(\Upsilon(2s))/\sigma(\Upsilon(3s))$ used in 5.02 TeV collisions. We obtain the differential cross section of $\Upsilon(3s)$ to be $d\sigma_{\Upsilon(3s)}/dy=3.58$ nb in the central rapidity. {For p-wave states, their cross sections at 2.76 TeV are obtained with the same ratio $\sigma(\Upsilon(1p,2p))/\sigma(\Upsilon(1s))$ used in 5.02 TeV.} The direct cross section before feed-down process can be obtained based on the same approach in the case of 5.02 TeV. The shadowing nuclear modification factor on bottomonium production is 
calculated with the EPS09 NLO model as done in 5.02 TeV. Hot medium evolution is also updated with a new input of the energy density profile, where the maximum local temperature at the center of the medium is determined as $T_0({\bf x}_T=0|b=0)=484$ MeV.  

In Fig.\ref{fig-2760RAA-Np}, the nuclear modification factor of three bottomonium 
states $\Upsilon(1s,2s,3s)$ are calculated with different complex potentials. The real part of the potential is taken as free energy and the internal 
energy $V_R=F$ and $U$. It seems that both potentials can explain the data of the ground and the excited state $\Upsilon(1s,2s)$. However, the relations between $R_{AA}(2s)$ and $R_{AA}(3S)$ become very different in the weak ($V_R=F$) and strong ($V_R=U$) binding scenarios. In the weak heavy-quark potential, bottomonium wave packages tend to expand outside in the hot medium. This results in the transition of $2s$ to $3s$ states, and gets $R_{AA}(3s)\ge R_{AA}(2s)$. 
When taking a 
strong potential $V_R=U$, theoretical studies show a clear pattern of sequential suppression between $\Upsilon(1s,2s,3s)$. More experimental data especially about the $\Upsilon(2s,3s)$ can help to reveal the in-medium heavy-quark potentials.

\begin{figure}[!tbp]
\centering 
\includegraphics[width=0.23\textwidth]{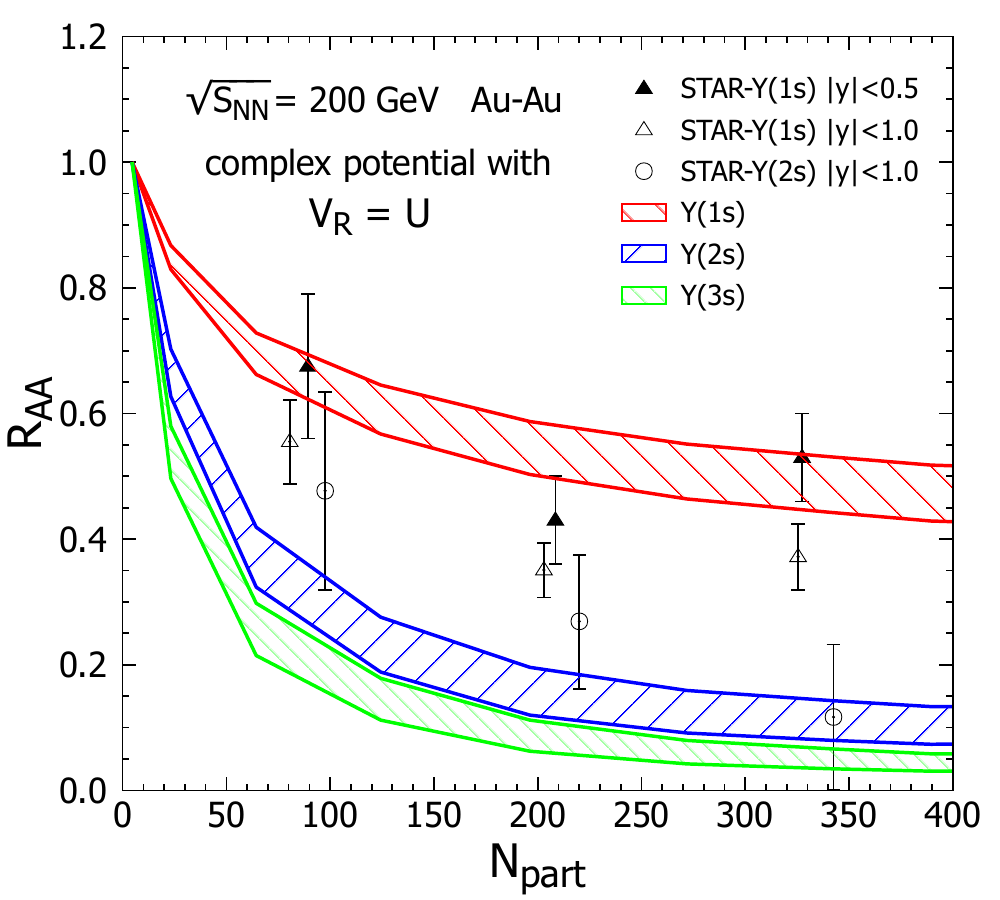}
\includegraphics[width=0.23\textwidth]{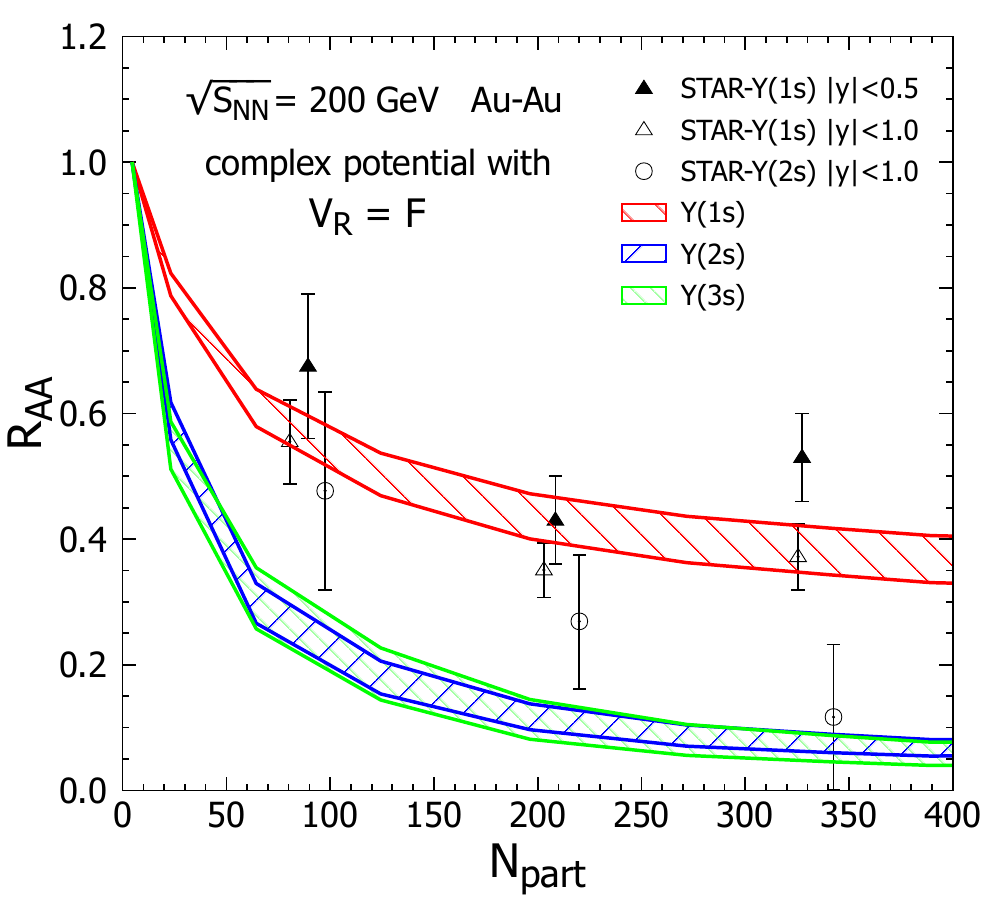}
\includegraphics[width=0.23\textwidth]{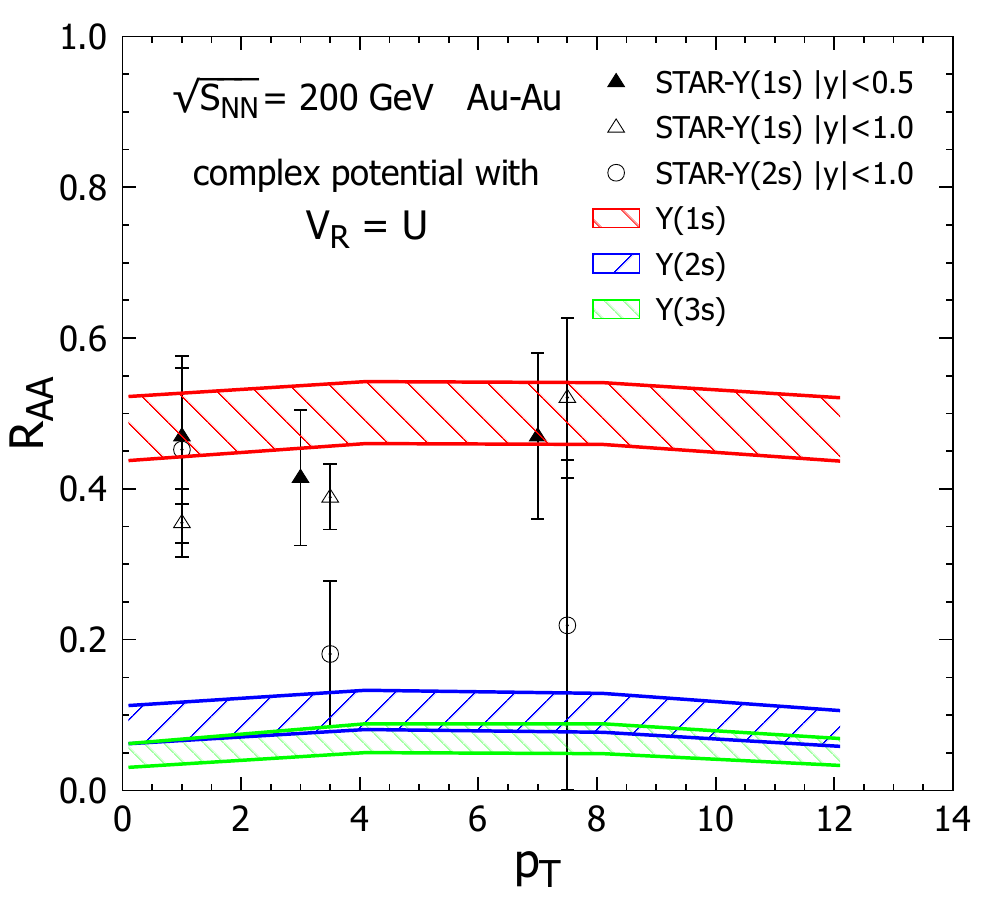}
\includegraphics[width=0.23\textwidth]{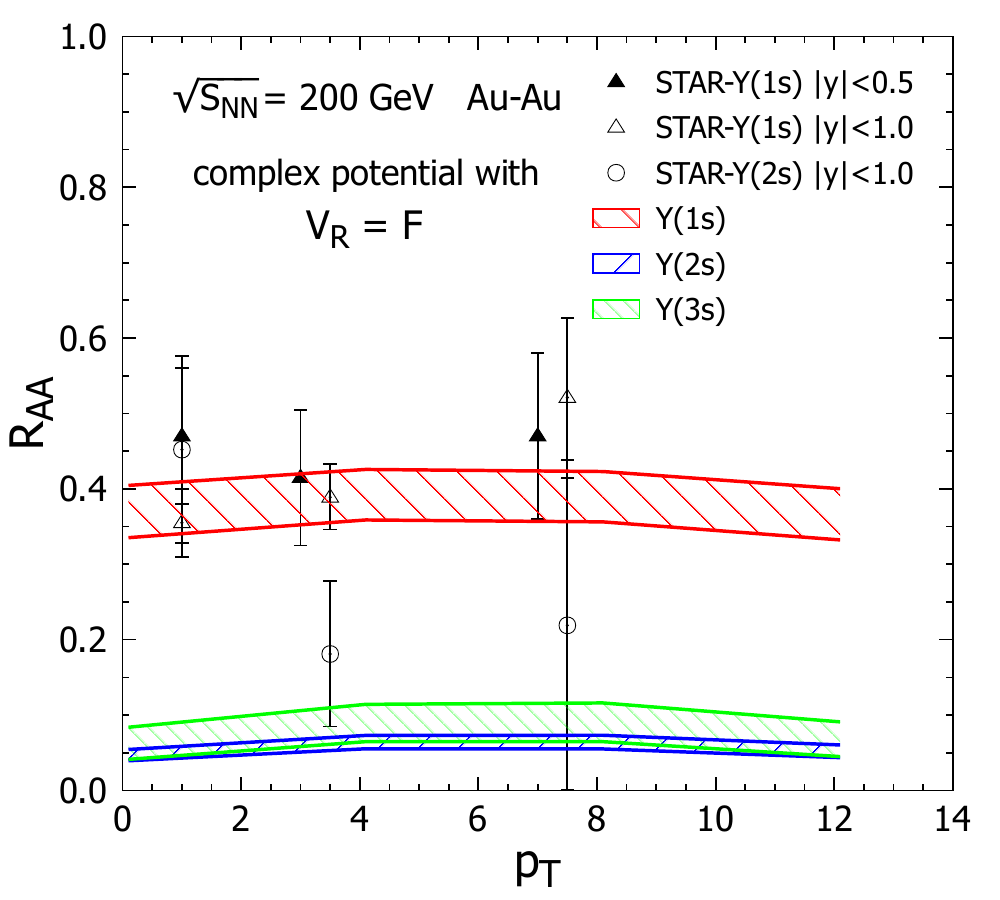}
\caption{\label{fig-200RAA-FU} 
The nuclear modification factors of bottomonium 
states $\Upsilon(1s,2s,3s)$ as a function of $N_p$ and $p_T$ in $\sqrt{s_{NN}}=200$ GeV Au-Au collisions. $R_{AA}(p_T)$ is in cent.0-60\%. 
The real part of the potential is taken as $V_R=F$ and $V_R=U$ respectively.  The band of the theoretical results corresponds to the uncertainty of the imaginary potential. 
The experimental data are cited from STAR Collaboration~\cite{Ye:2017fwv,STAR:2022rpk}.
}
\end{figure}

At the end, we turn to the bottomonium suppression at $\sqrt{s_{NN}}=200$ GeV Au-Au collisions. In the central rapidity 
of pp collision, bottomonium different 
cross section is fitted to be $d\sigma_{\Upsilon(1s,2s,3s)}/dy=(2.35,0.77,0.27)$ nb based on the $\Upsilon(1s,2s)$ measurements from STAR Collaboration, while $\Upsilon(3s)$ cross section is scaled with the same ratio of $\sigma(\Upsilon(2s))/\sigma(\Upsilon(3s))$ as in 5.02 TeV.  The medium temperatures at RHIC energy is much lower than the values at LHC energies. The maximum temperature 
in the central Au-Au collisions is determined to be $T_0({\bf x}_T=0|b=0)=390$ MeV. The shadowing factor on bottomonium is much smaller at RHIC energy. With weak and strong potentials, we calculated the $R_{AA}(1s,2s,3s)$ as a function of $N_p$ and $p_T$ respectively in Fig.\ref{fig-200RAA-FU}. Even theoretical bands with different choices of $V_R$ can explain the $\Upsilon(1s)$ data, but the pattern of sequential suppression in $\Upsilon(2s,3s)$ can only be seen in the 
situation of $V_R=U$. Employing different imaginary potentials do not change the suppression pattern in three states $\Upsilon(1s,2s,3s)$. 
{For excited states, theoretical 
bands slightly underestimate the experimental data but 
still stay in the uncertainty of the data points. 
This is 
due to the imaginary potential which results in strong 
suppression on excited states in the temperature regions at RHIC energy.}

\section{Uncertainty discussion}

\begin{figure}[tbp]
\centering 
\includegraphics[width=0.23\textwidth]{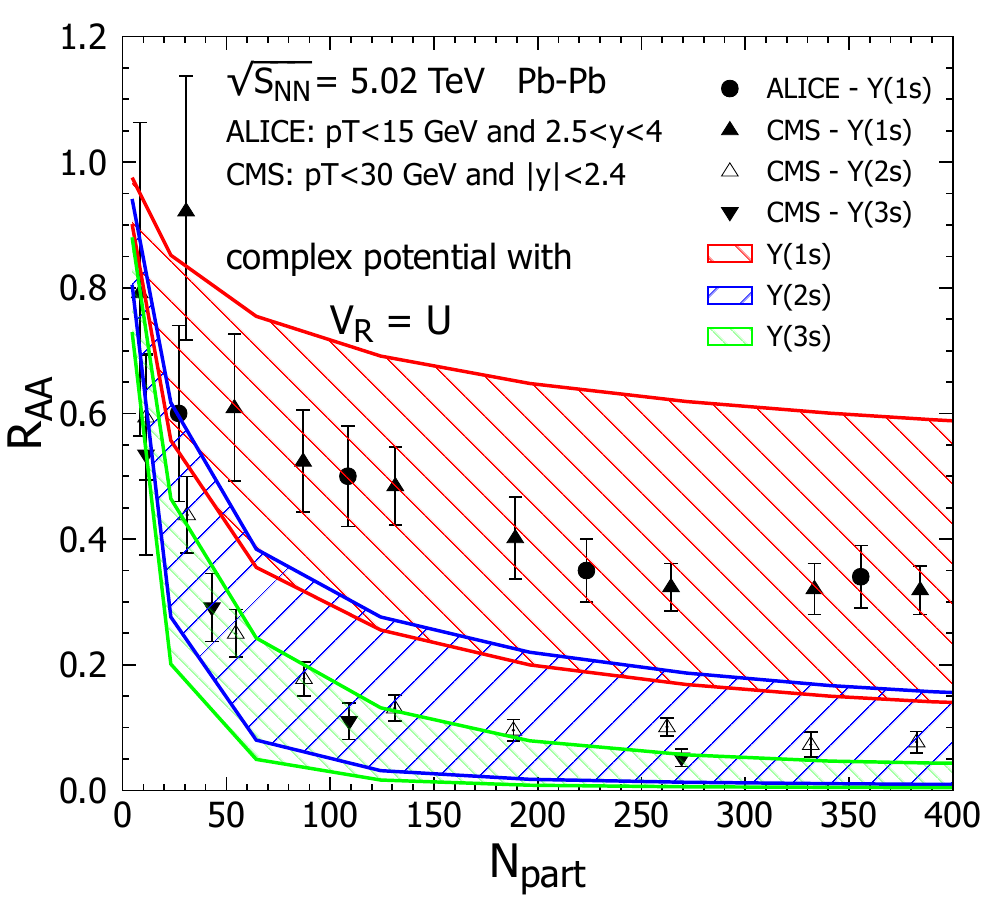}
\includegraphics[width=0.23\textwidth]{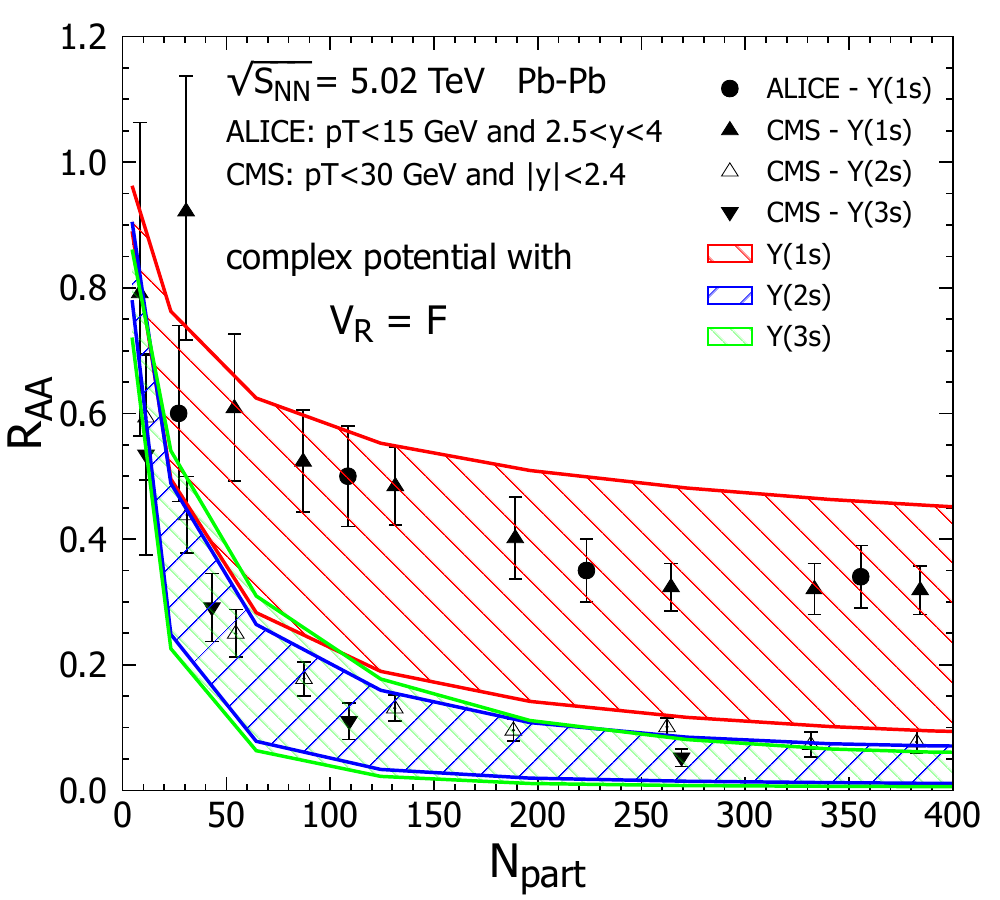}
\includegraphics[width=0.23\textwidth]{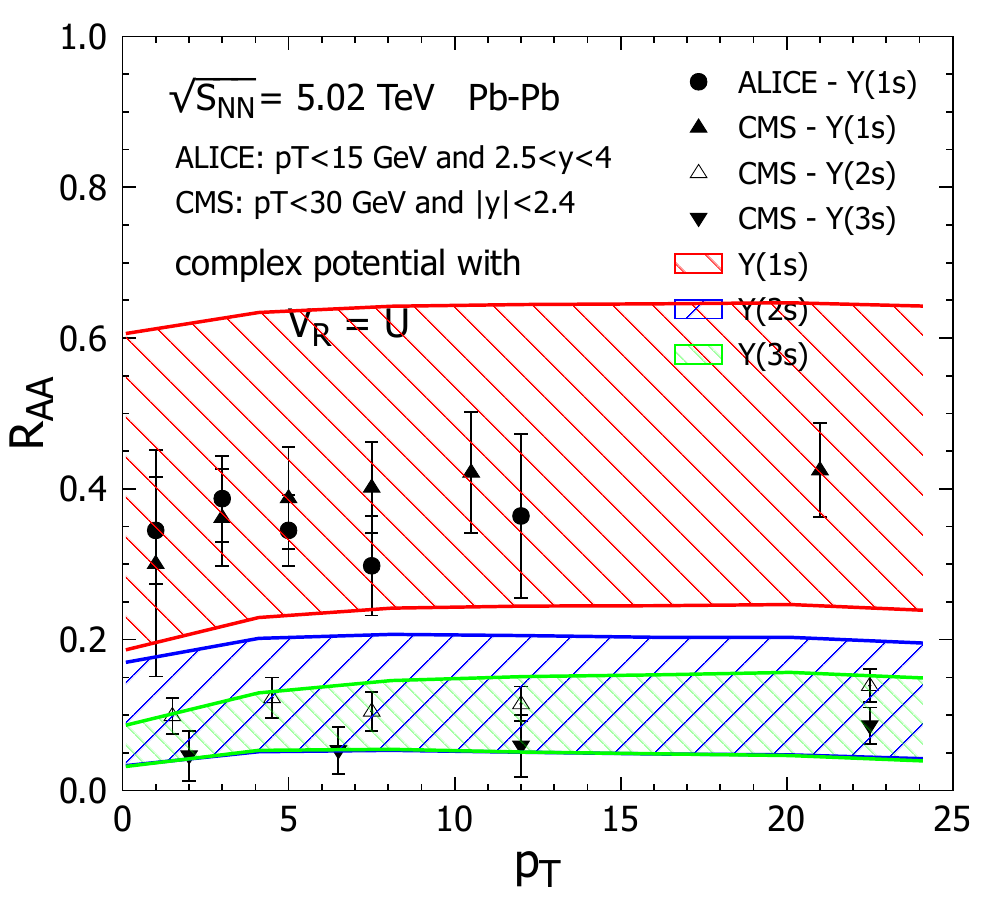}
\includegraphics[width=0.23\textwidth]{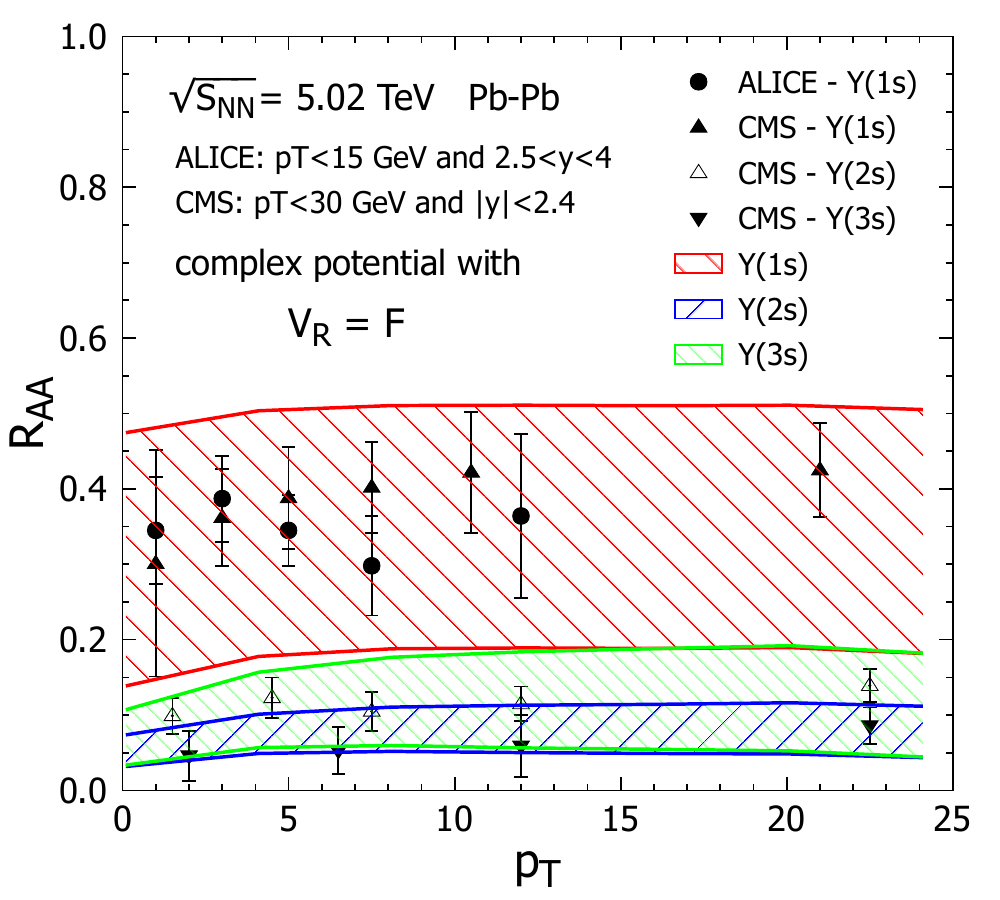}
\caption{\label{fig-RAA-5TeV-largeVI} 
The nuclear modification factors of bottomonium 
$\Upsilon(1s,2s,3s)$ in $\sqrt{s_{NN}}=5.02$ TeV Pb-Pb collisions. The imaginary potential labeled with ``Band 2'' in Fig.\ref{fig:ImagV} is taken. Other parameters are the same as in the previous figures. 
}
\end{figure}

\begin{figure}[tbp]
\centering 
\includegraphics[width=0.23\textwidth]{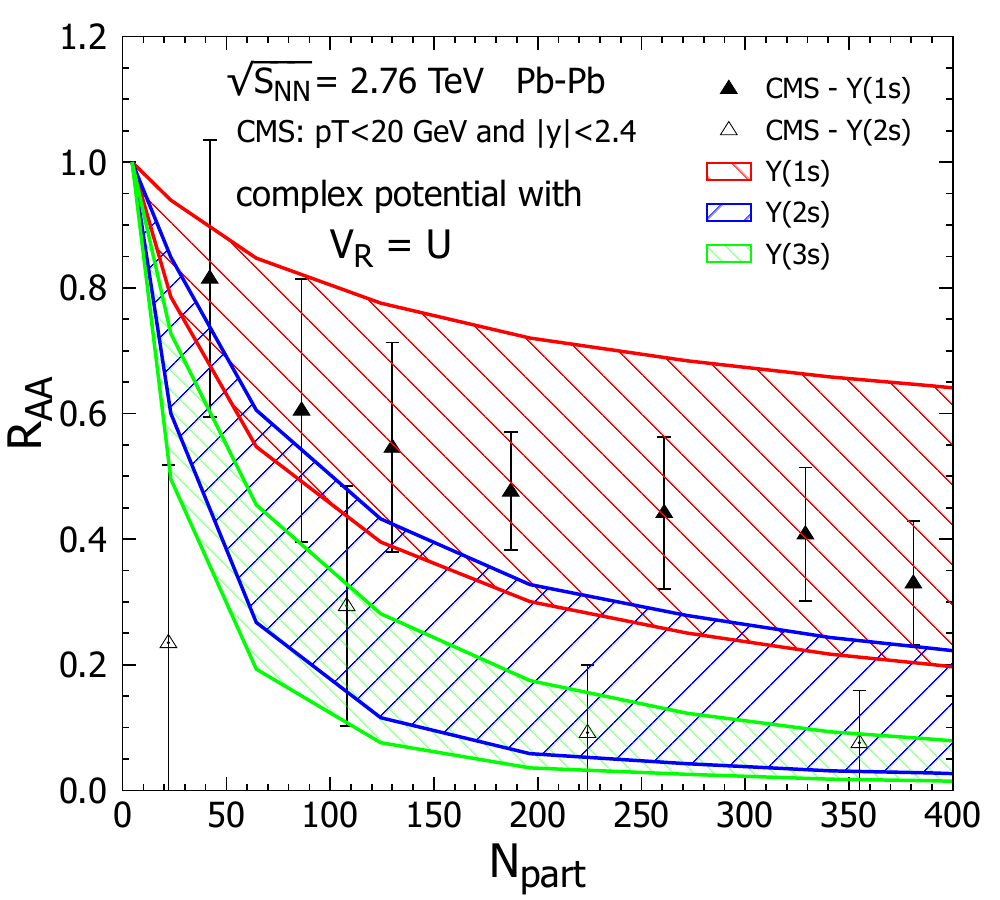}
\includegraphics[width=0.23\textwidth]{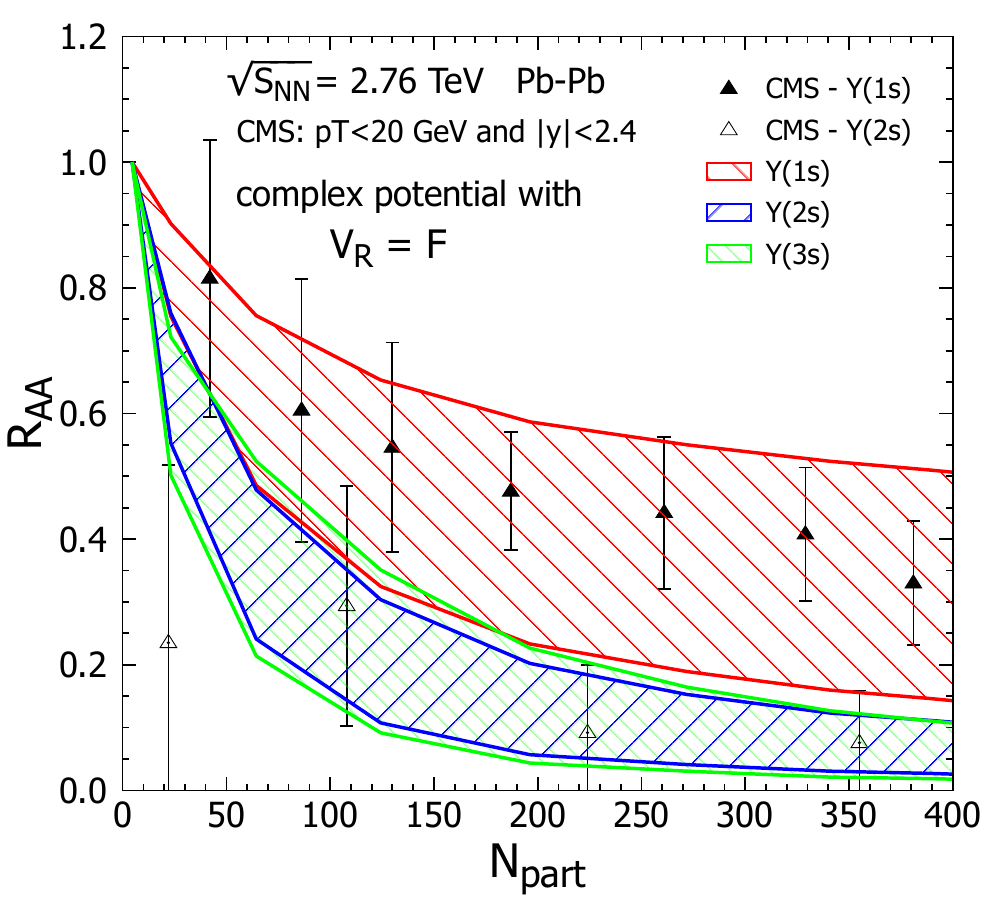}
\includegraphics[width=0.23\textwidth]{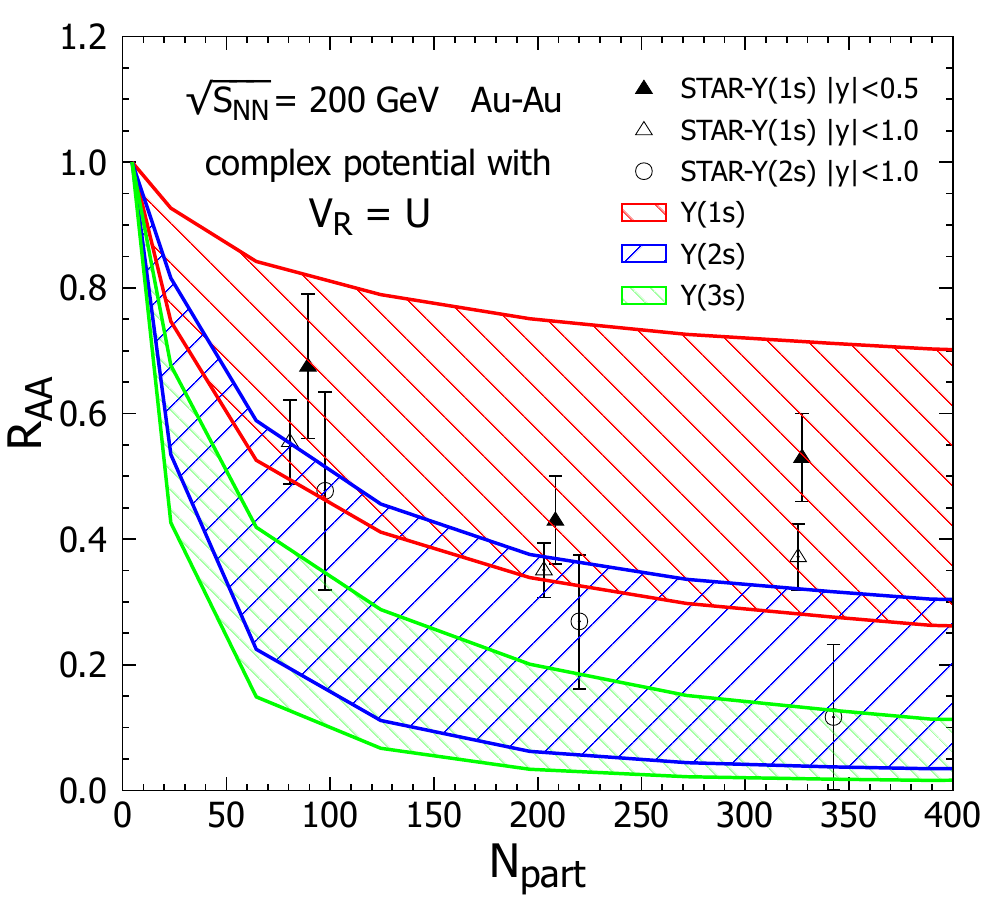}
\includegraphics[width=0.23\textwidth]{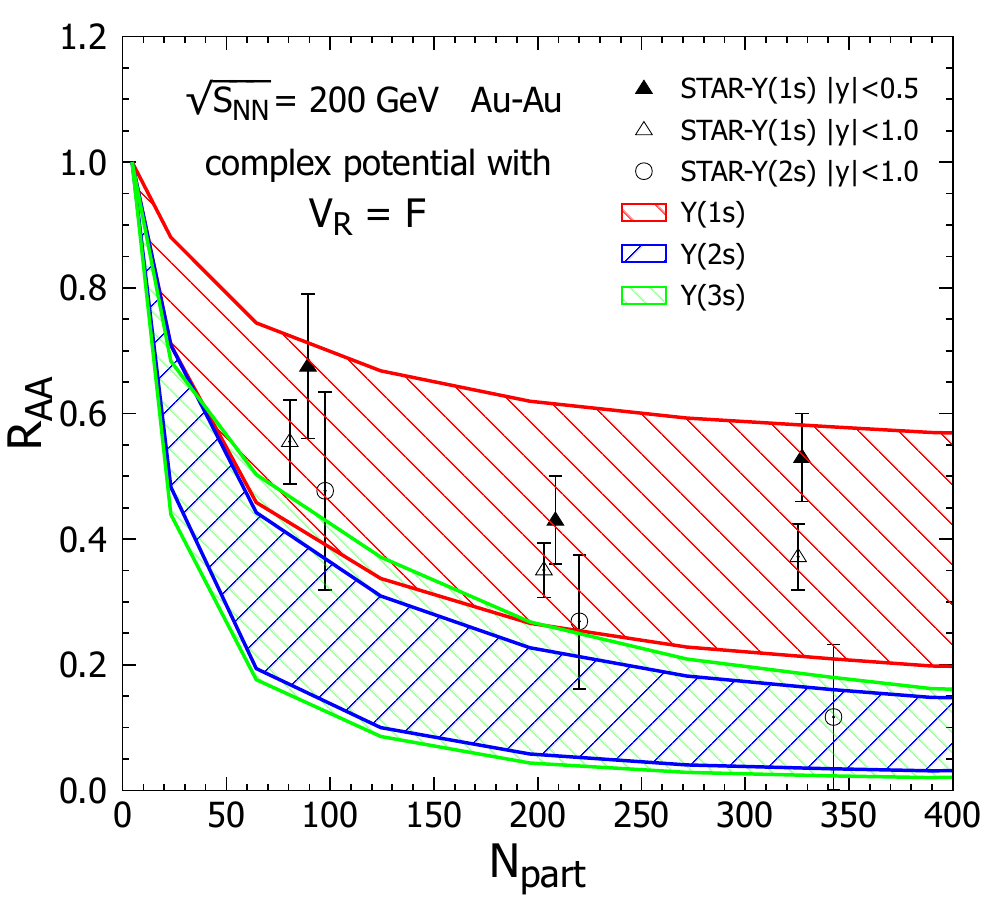}
\includegraphics[width=0.23\textwidth]{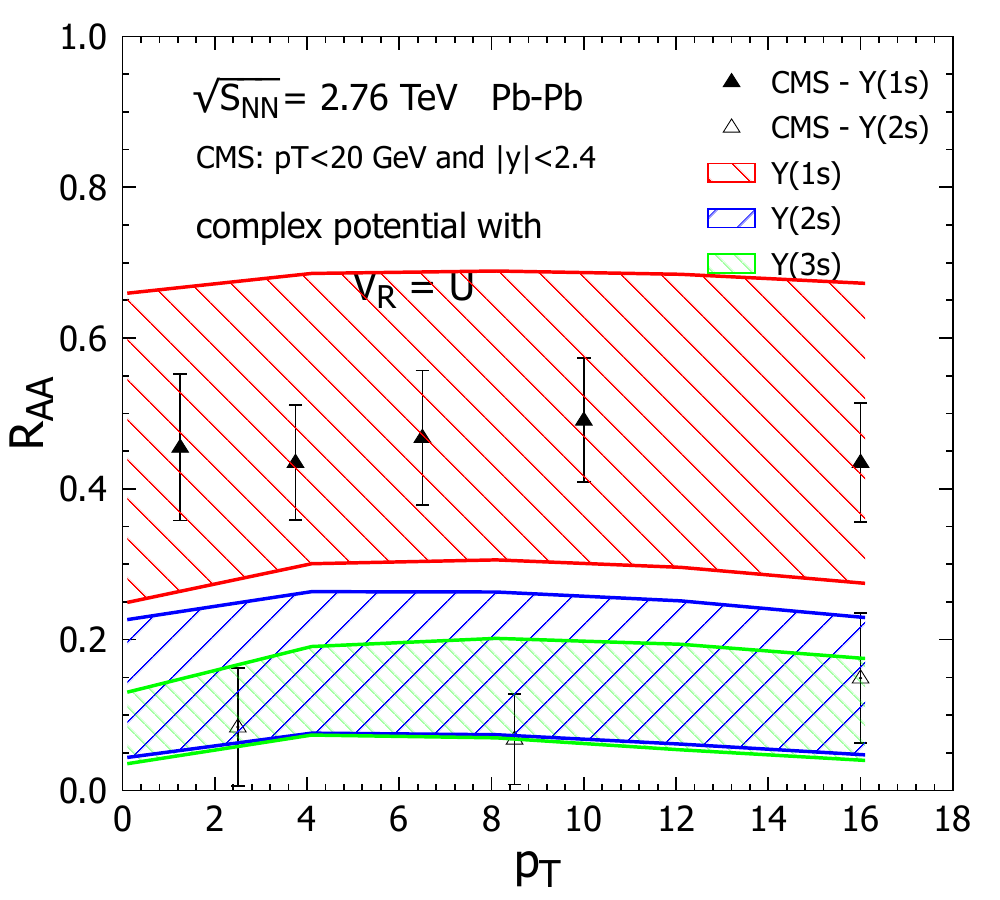}
\includegraphics[width=0.23\textwidth]{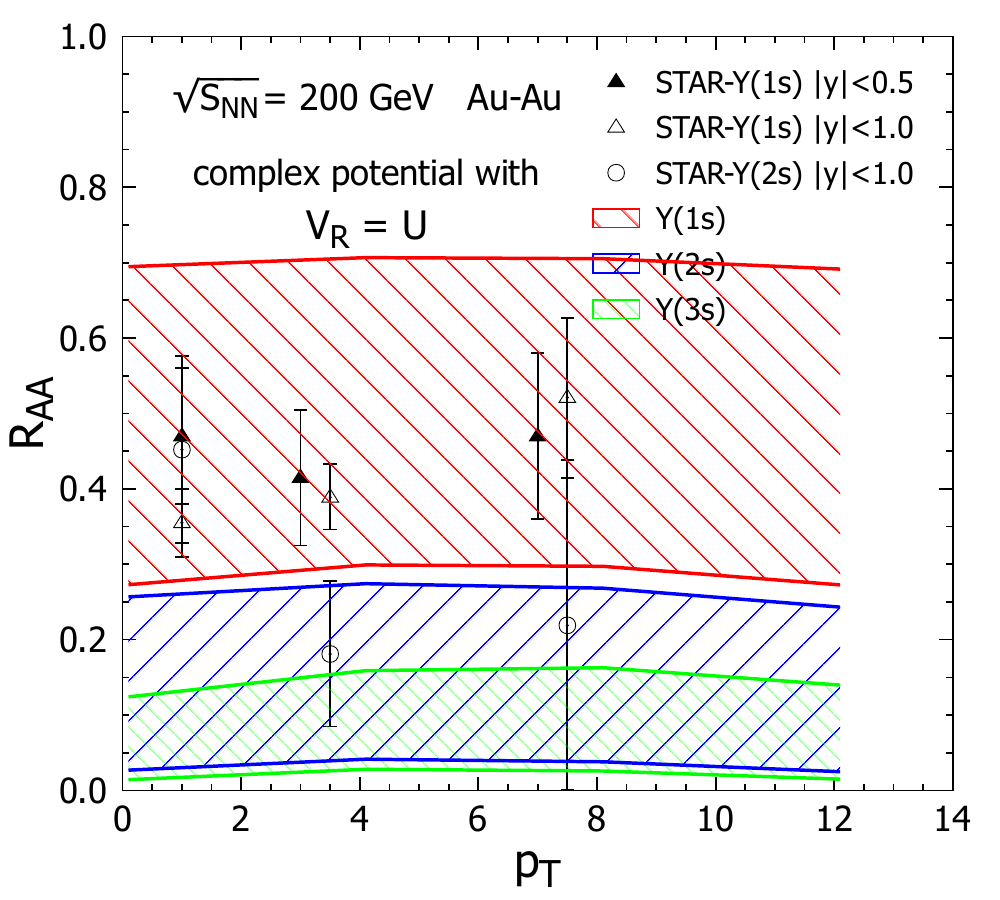}
\caption{\label{fig-RAA-2760-200-largeVI} 
The nuclear modification factors of bottomonium 
states $\Upsilon(1s,2s,3s)$ in $\sqrt{s_{NN}}=2.76$ TeV Pb-Pb and 200 GeV Au-Au collisions. The imaginary potential labeled with ``Band 2'' in Fig.\ref{fig:ImagV} is taken. Other parameters are the same as in the previous figures. 
}
\end{figure}

Now we discuss the sensitivity of the results on the parameters. As final bottomonium is dominated by the primordial production, the initial $p_T$
distribution function of bottomonium Eq.(\ref{eq:pp-input}) is mostly cancelled in the numerator and denominator of the $R_{AA}(p_T)$. Besides, $R_{AA}$ shows weak dependence on $p_T$ in a fixed centrality. Therefore, $R_{AA}$ depends weakly on the choices of $n$ and $\langle p_T^2\rangle_{pp}$ in the initial $p_T$ distribution. The rapidity differential cross section $d\sigma/dy$ used in both numerator and denominator of $R_{AA}$ is also canceled. Another important ingredient is the imaginary potential. The degree of bottomonium suppression depends on the magnitude of the imaginary potential. As imaginary potential increases with the radius, bottomonium excited states suffer stronger suppression compared with the ground state. In previous sections, bottomonium $R_{AA}$ are calculated by taking the $V_I$ labeled with ``Band 1''. When we take $V_I$ with one standard deviation error bar (``Band 2''), the corresponding $R_{AA}$s are also calculated and plotted in Fig.\ref{fig-RAA-5TeV-largeVI}-\ref{fig-RAA-2760-200-largeVI}. As one can see, the uncertainty in $R_{AA}$ becomes very large to prevent any solid conclusions. While the sequential suppression pattern of $\Upsilon(1s,2s,3s)$ can still be seen when we take a strong heavy quark potential. That means the conclusion of this work is not changed by different choices of the imaginary potential.

\section{Summary}

In this work, we employ the time-dependent Schr\"odinger equation with different complex potentials to study the sequential suppression of bottomonium states $\Upsilon(1s,2s,3s)$ at $\sqrt{s_{NN}}=(5.02, 2.76)$ TeV Pb-Pb collisions and 200 GeV Au-Au  collisions. The imaginary part of the heavy quark potential is fitted with the lattice QCD calculations, where the uncertainty of the data points are considered in the imaginary potential with a band. While the real part of the potential is  between the free energy $F$ and the internal energy $U$. The realistic real potential is 
parametrized as a function of $F$ and $U$. 
With different real potentials, the attracted force in the wave packages of bottomonium becomes  different. We find that the sequential suppression pattern of $\Upsilon(1s,2s,3s)$ is explained well when the real potential is strong and close to $U$ which can constrain the wave package in the hot medium. This conclusion is not changed by different choices of the imaginary potential. Therefore, the sequential suppression pattern of bottomonium states is suggested to be a probe of the strong in-medium heavy quark potential in heavy-ion collisions.

\vspace{2cm}
{\bf Acknowledgement:}
This work is supported by the National Natural Science Foundation of China
(NSFC) under Grant Nos. 12175165.

%\end{spacing}

\vspace{2cm}
\begin{thebibliography}{20}

\bibitem{Bazavov:2011nk}
A.~Bazavov, T.~Bhattacharya, M.~Cheng, C.~DeTar, H.~T.~Ding, S.~Gottlieb, R.~Gupta, P.~Hegde, U.~M.~Heller and F.~Karsch, \textit{et al.}
%``The chiral and deconfinement aspects of the QCD transition,''
Phys. Rev. D \textbf{85}, 054503 (2012)
%doi:10.1103/PhysRevD.85.054503
%[arXiv:1111.1710 [hep-lat]].

\bibitem{Matsui:1986dk}
T.~Matsui and H.~Satz,
%``$J/\psi$ Suppression by Quark-Gluon Plasma Formation,''
Phys. Lett. B \textbf{178}, 416-422 (1986)
%doi:10.1016/0370-2693(86)91404-8

\bibitem{Satz:2005hx}
H.~Satz,
%``Colour deconfinement and quarkonium binding,''
J. Phys. G \textbf{32}, R25 (2006)
%doi:10.1088/0954-3899/32/3/R01
%[arXiv:hep-ph/0512217 [hep-ph]].

\bibitem{Zhao:2020jqu}
J.~Zhao, K.~Zhou, S.~Chen and P.~Zhuang,
%``Heavy flavors under extreme conditions in high energy nuclear collisions,''
Prog. Part. Nucl. Phys. \textbf{114}, 103801 (2020)
%doi:10.1016/j.ppnp.2020.103801
%[arXiv:2005.08277 [nucl-th]].

\bibitem{Chen:2019qzx}
B.~Chen, M.~Hu, H.~Zhang and J.~Zhao,
%``Probe the tilted Quark-Gluon Plasma with charmonium directed flow,''
Phys. Lett. B \textbf{802}, 135271 (2020)
%doi:10.1016/j.physletb.2020.135271
%[arXiv:1910.08275 [nucl-th]].

\bibitem{Zhao:2021voa}
J.~Zhao, B.~Chen and P.~Zhuang,
%``Charmonium triangular flow in high energy nuclear collisions,''
Phys. Rev. C \textbf{105}, no.3, 034902 (2022)
%doi:10.1103/PhysRevC.105.034902
%[arXiv:2112.00293 [hep-ph]].

\bibitem{Liu:2010ej}
Y.~Liu, B.~Chen, N.~Xu and P.~Zhuang,
%``$\Upsilon$ Production as a Probe for Early State Dynamics in High Energy Nuclear Collisions at RHIC,''
Phys. Lett. B \textbf{697}, 32-36 (2011)
%doi:10.1016/j.physletb.2011.01.026
%[arXiv:1009.2585 [nucl-th]].



\bibitem{Du:2018wsj}
X.~Du and R.~Rapp,
%``In-Medium Charmonium Production in Proton-Nucleus Collisions,''
JHEP \textbf{03}, 015 (2019)
%doi:10.1007/JHEP03(2019)015
%[arXiv:1808.10014 [nucl-th]].

\bibitem{Strickland:2011mw}
M.~Strickland,
%``Thermal $\upsilon_{1s}$ and chi\_b1 suppression in $\sqrt{s_{NN}}=2.76$ TeV Pb-Pb collisions at the LHC,''
Phys. Rev. Lett. \textbf{107}, 132301 (2011)
%doi:10.1103/PhysRevLett.107.132301
%[arXiv:1106.2571 [hep-ph]].

\bibitem{Strickland:2011aa}
M.~Strickland and D.~Bazow,
%``Thermal Bottomonium Suppression at RHIC and LHC,''
Nucl. Phys. A \textbf{879}, 25-58 (2012)
%doi:10.1016/j.nuclphysa.2012.02.003
%[arXiv:1112.2761 [nucl-th]].

\bibitem{Brambilla:2021wkt}
N.~Brambilla, M.~\'A.~Escobedo, M.~Strickland, A.~Vairo, P.~Vander Griend and J.~H.~Weber,
%``Bottomonium production in heavy-ion collisions using quantum trajectories: Differential observables and momentum anisotropy,''
Phys. Rev. D \textbf{104}, no.9, 094049 (2021)
%doi:10.1103/PhysRevD.104.094049
%[arXiv:2107.06222 [hep-ph]].


\bibitem{Yao:2021lus}
X.~Yao,
%``Open quantum systems for quarkonia,''
Int. J. Mod. Phys. A \textbf{36}, no.20, 2130010 (2021)
%doi:10.1142/S0217751X21300106
%[arXiv:2102.01736 [hep-ph]].

\bibitem{Yan:2006ve}
L.~Yan, P.~Zhuang and N.~Xu,
%``Competition between J / psi suppression and regeneration in quark-gluon plasma,''
Phys. Rev. Lett. \textbf{97}, 232301 (2006)
%doi:10.1103/PhysRevLett.97.232301
%[arXiv:nucl-th/0608010 [nucl-th]].

\bibitem{Chen:2016dke}
B.~Chen, T.~Guo, Y.~Liu and P.~Zhuang,
%``Cold and Hot Nuclear Matter Effects on Charmonium Production in p+Pb Collisions at LHC Energy,''
Phys. Lett. B \textbf{765}, 323-327 (2017)
%doi:10.1016/j.physletb.2016.12.021
%[arXiv:1607.07927 [nucl-th]].

\bibitem{Zhao:2007hh}
X.~Zhao and R.~Rapp,
%``Transverse Momentum Spectra of $J/\psi$ in Heavy-Ion Collisions,''
Phys. Lett. B \textbf{664}, 253-257 (2008)
%doi:10.1016/j.physletb.2008.03.068
%[arXiv:0712.2407 [hep-ph]].

\bibitem{Du:2015wha}
X.~Du and R.~Rapp,
%``Sequential Regeneration of Charmonia in Heavy-Ion Collisions,''
Nucl. Phys. A \textbf{943}, 147-158 (2015)
%doi:10.1016/j.nuclphysa.2015.09.006
%[arXiv:1504.00670 [hep-ph]].

\bibitem{Yao:2020xzw}
X.~Yao, W.~Ke, Y.~Xu, S.~A.~Bass and B.~M\"uller,
%``Coupled Boltzmann Transport Equations of Heavy Quarks and Quarkonia in Quark-Gluon Plasma,''
JHEP \textbf{01}, 046 (2021)
%doi:10.1007/JHEP01(2021)046
%[arXiv:2004.06746 [hep-ph]].

\bibitem{Yao:2020eqy}
X.~Yao and T.~Mehen,
%``Quarkonium Semiclassical Transport in Quark-Gluon Plasma: Factorization and Quantum Correction,''
JHEP \textbf{02}, 062 (2021)
%doi:10.1007/JHEP02(2021)062
%[arXiv:2009.02408 [hep-ph]].

\bibitem{Zhao:2022ggw}
J.~Zhao and P.~Zhuang,
Phys. Rev. C \textbf{105}, no.6, 064907 (2022)
%doi:10.1103/PhysRevC.105.064907


\bibitem{Blaizot:2018oev}
J.~P.~Blaizot and M.~A.~Escobedo,
%``Approach to equilibrium of a quarkonium in a quark-gluon plasma,''
Phys. Rev. D \textbf{98}, no.7, 074007 (2018)
%doi:10.1103/PhysRevD.98.074007
%[arXiv:1803.07996 [hep-ph]].
\bibitem{Blaizot:2021xqa}
J.~P.~Blaizot and M.~\'A.~Escobedo,
%``Phenomenological study of quarkonium suppression and the impact of the energy gap between singlets and octets,''
Phys. Rev. D \textbf{104}, no.5, 054034 (2021)
%doi:10.1103/PhysRevD.104.054034
%[arXiv:2106.15371 [hep-ph]].

\bibitem{Katz:2015qja}
R.~Katz and P.~B.~Gossiaux,
%``The Schr\"odinger\textendash{}Langevin equation with and without thermal fluctuations,''
Annals Phys. \textbf{368}, 267-295 (2016)
%doi:10.1016/j.aop.2016.02.005
%[arXiv:1504.08087 [quant-ph]].



\bibitem{Gossiaux:2016htk}
P.~B.~Gossiaux and R.~Katz,
%``Upsilon suppression in the Schr\"odinger\textendash{}Langevin approach,''
Nucl. Phys. A \textbf{956}, 737-740 (2016)
%doi:10.1016/j.nuclphysa.2016.04.017
%[arXiv:1601.01443 [hep-ph]].

\bibitem{Akamatsu:2014qsa}
Y.~Akamatsu,
%``Heavy quark master equations in the Lindblad form at high temperatures,''
Phys. Rev. D \textbf{91}, no.5, 056002 (2015)
%doi:10.1103/PhysRevD.91.056002
%[arXiv:1403.5783 [hep-ph]].

\bibitem{Brambilla:2016wgg}
N.~Brambilla, M.~A.~Escobedo, J.~Soto and A.~Vairo,
%``Quarkonium suppression in heavy-ion collisions: an open quantum system approach,''
Phys. Rev. D \textbf{96}, no.3, 034021 (2017)
%doi:10.1103/PhysRevD.96.034021
%[arXiv:1612.07248 [hep-ph]].

\bibitem{Brambilla:2020qwo}
N.~Brambilla, M.~\'A.~Escobedo, M.~Strickland, A.~Vairo, P.~Vander Griend and J.~H.~Weber,
%``Bottomonium suppression in an open quantum system using the quantum trajectories method,''
JHEP \textbf{05}, 136 (2021)
%doi:10.1007/JHEP05(2021)136
%[arXiv:2012.01240 [hep-ph]].

\bibitem{Akamatsu:2018xim}
Y.~Akamatsu, M.~Asakawa, S.~Kajimoto and A.~Rothkopf,
%``Quantum dissipation of a heavy quark from a nonlinear stochastic Schr\"odinger equation,''
JHEP \textbf{07}, 029 (2018)
%doi:10.1007/JHEP07(2018)029
%[arXiv:1805.00167 [nucl-th]].

\bibitem{Xie:2022tzs}
Z.~Xie and B.~Chen,
%``Open quantum system approach for heavy quark thermalization,''
[arXiv:2205.13302 [nucl-th]].

\bibitem{CMS:2018zza}
A.~M.~Sirunyan \textit{et al.} [CMS],
%``Measurement of nuclear modification factors of $\Upsilon$(1S), $\Upsilon$(2S), and $\Upsilon$(3S) mesons in PbPb collisions at $\sqrt{s_{_\mathrm{NN}}} =$ 5.02 TeV,''
Phys. Lett. B \textbf{790}, 270-293 (2019)
%doi:10.1016/j.physletb.2019.01.006
%[arXiv:1805.09215 [hep-ex]].

\bibitem{CMS:2016rpc}
V.~Khachatryan \textit{et al.} [CMS],
%``Suppression of $\Upsilon(1S), \Upsilon(2S)$ and $\Upsilon(3S)$ production in PbPb collisions at $\sqrt{s_{\rm NN}}$ = 2.76 TeV,''
Phys. Lett. B \textbf{770}, 357-379 (2017)
%doi:10.1016/j.physletb.2017.04.031
%[arXiv:1611.01510 [nucl-ex]].


\bibitem{ALICE:2014wnc}
B.~B.~Abelev \textit{et al.} [ALICE],
%``Suppression of $\Upsilon (1S)$ at forward rapidity in Pb-Pb collisions at $\sqrt{s_{\rm NN}} = 2.76$ TeV,''
Phys. Lett. B \textbf{738}, 361-372 (2014)
%doi:10.1016/j.physletb.2014.10.001
%[arXiv:1405.4493 [nucl-ex]].

\bibitem{PHENIX:2014tbe}
A.~Adare \textit{et al.} [PHENIX],
%``Measurement of $\Upsilon(1S+2S+3S)$ production in $p+p$ and Au$+$Au collisions at $\sqrt{s_{_{NN}}}=200$ GeV,''
Phys. Rev. C \textbf{91}, no.2, 024913 (2015)
%doi:10.1103/PhysRevC.91.024913
%[arXiv:1404.2246 [nucl-ex]].

\bibitem{STAR:2013kwk}
L.~Adamczyk \textit{et al.} [STAR],
%``Suppression of $\Upsilon$ production in d+Au and Au+Au collisions at $\sqrt{s_{NN}}$=200 GeV,''
Phys. Lett. B \textbf{735}, 127-137 (2014)
[erratum: Phys. Lett. B \textbf{743}, 537-541 (2015)]
%doi:10.1016/j.physletb.2014.06.028
%[arXiv:1312.3675 [nucl-ex]].

\bibitem{Chen:2017duy}
B.~Chen and J.~Zhao,
%``Bottomonium Continuous Production from Unequilibrium Bottom Quarks in Ultrarelativistic Heavy Ion Collisions,''
Phys. Lett. B \textbf{772}, 819-824 (2017)
%doi:10.1016/j.physletb.2017.07.054
%[arXiv:1704.05622 [nucl-th]].
\bibitem{Blaizot:2015hya}
J.~P.~Blaizot, D.~De Boni, P.~Faccioli and G.~Garberoglio,
%``Heavy quark bound states in a quark\textendash{}gluon plasma: Dissociation and recombination,''
Nucl. Phys. A \textbf{946}, 49-88 (2016)
%doi:10.1016/j.nuclphysa.2015.10.011
%[arXiv:1503.03857 [nucl-th]].


\bibitem{Wen:2022utn}
L. Wen, X. Du, S. Shi, B. Chen, 
%``Probe the color screening in proton-nucleus collisions with complex potentials,''
Chin.Phys. C 46,(2022) 114102, 
% https://doi.org/10.1088/1674-1137/ac7fe6 
%[arXiv:2205.07520 [nucl-th]].


\bibitem{Du:2019tjf}
X.~Du, S.~Y.~F.~Liu and R.~Rapp,
%``Extraction of the Heavy-Quark Potential from Bottomonium Observables in Heavy-Ion Collisions,''
Phys. Lett. B \textbf{796}, 20-25 (2019)
%doi:10.1016/j.physletb.2019.07.032
%[arXiv:1904.00113 [nucl-th]].

\bibitem{Islam:2020bnp}
A.~Islam and M.~Strickland,
%``Bottomonium suppression and elliptic flow using Heavy Quarkonium Quantum Dynamics,''
JHEP \textbf{21}, 235 (2020)
%doi:10.1007/JHEP03(2021)235
%[arXiv:2010.05457 [hep-ph]].

\bibitem{Burnier:2016mxc}
Y.~Burnier and A.~Rothkopf,
%``Complex heavy-quark potential and Debye mass in a gluonic medium from lattice QCD,''
Phys. Rev. D \textbf{95}, no.5, 054511 (2017)
%doi:10.1103/PhysRevD.95.054511
%[arXiv:1607.04049 [hep-lat]].

\bibitem{Islam:2020gdv}
A.~Islam and M.~Strickland,
%``Bottomonium suppression and elliptic flow from real-time quantum evolution,''
Phys. Lett. B \textbf{811}, 135949 (2020)
%doi:10.1016/j.physletb.2020.135949
%[arXiv:2007.10211 [hep-ph]].

\bibitem{Shi:2021qri}
S.~Shi, K.~Zhou, J.~Zhao, S.~Mukherjee and P.~Zhuang,
%``Heavy quark potential in the quark-gluon plasma: Deep neural network meets lattice quantum chromodynamics,''
Phys. Rev. D \textbf{105}, no.1, 1 (2022)
%doi:10.1103/PhysRevD.105.014017
%[arXiv:2105.07862 [hep-ph]].

\bibitem{Zhao:2017yhj}
W.~Zhao, H.~j.~Xu and H.~Song,
%``Collective flow in 2.76 A TeV and 5.02 A TeV Pb+Pb collisions,''
Eur. Phys. J. C \textbf{77}, no.9, 645 (2017)
%doi:10.1140/epjc/s10052-017-5186-x
%[arXiv:1703.10792 [nucl-th]].

\bibitem{ParticleDataGroup:2018ovx}
M.~Tanabashi \textit{et al.} [Particle Data Group],
%``Review of Particle Physics,''
Phys. Rev. D \textbf{98}, no.3, 030001 (2018)
%doi:10.1103/PhysRevD.98.030001

%\bibitem{CMS:2016rpc}
%V.~Khachatryan \textit{et al.} [CMS],
%%``Suppression of $\Upsilon(1S), \Upsilon(2S)$ and $\Upsilon(3S)$ production in PbPb collisions at $\sqrt{s_{\rm NN}}$ = 2.76 TeV,''
%Phys. Lett. B \textbf{770}, 357-379 (2017)
%doi:10.1016/j.physletb.2017.04.031
%[arXiv:1611.01510 [nucl-ex]].

\bibitem{CMS:2013qur}
S.~Chatrchyan \textit{et al.} [CMS],
%``Measurement of the $\Upsilon(1S), \Upsilon(2S)$, and $\Upsilon(3S)$ Cross Sections in $pp$ Collisions at $\sqrt{s}$ = 7 TeV,''
Phys. Lett. B \textbf{727}, 101-125 (2013)
%doi:10.1016/j.physletb.2013.10.033
%[arXiv:1303.5900 [hep-ex]].

\bibitem{ATLAS:2012lmu}
G.~Aad \textit{et al.} [ATLAS],
%``Measurement of Upsilon production in 7 TeV pp collisions at ATLAS,''
Phys. Rev. D \textbf{87}, no.5, 052004 (2013)
%doi:10.1103/PhysRevD.87.052004
%[arXiv:1211.7255 [hep-ex]].

\bibitem{CMS:2010wld}
V.~Khachatryan \textit{et al.} [CMS],
%``Upsilon Production Cross-Section in pp Collisions at $\sqrt{s}$=7 TeV,''
Phys. Rev. D \textbf{83}, 112004 (2011)
%doi:10.1103/PhysRevD.83.112004
%[arXiv:1012.5545 [hep-ex]].

\bibitem{LHCb:2012aa}
R.~Aaij \textit{et al.} [LHCb],
%``Measurement of Upsilon production in pp collisions at $\sqrt{s}$ = 7 TeV,''
Eur. Phys. J. C \textbf{72}, 2025 (2012)
%doi:10.1140/epjc/s10052-012-2025-y
%[arXiv:1202.6579 [hep-ex]].

\bibitem{LHCb:2014dei}
R.~Aaij \textit{et al.} [LHCb],
%``Measurement of $\Upsilon$ production in $pp$ collisions at $\sqrt{s}=2.76$ TeV,''
Eur. Phys. J. C \textbf{74}, no.4, 2835 (2014)
%doi:10.1140/epjc/s10052-014-2835-1
%[arXiv:1402.2539 [hep-ex]].

\bibitem{LHCb:2014ngh}
R.~Aaij \textit{et al.} [LHCb],
%``Study of $\chi _{{\mathrm {b}}}$ meson production in $\mathrm {p} $ $\mathrm {p} $ collisions at $\sqrt{s}=7$ and $8{\mathrm {\,TeV}} $ and observation of the decay $\chi _{{\mathrm {b}}}\mathrm {(3P)} \rightarrow \Upsilon \mathrm {(3S)} {\gamma } $,''
Eur. Phys. J. C \textbf{74}, no.10, 3092 (2014)
%doi:10.1140/epjc/s10052-014-3092-z
%[arXiv:1407.7734 [hep-ex]].

\bibitem{ALICE:2015pgg}
J.~Adam \textit{et al.} [ALICE],
%``Inclusive quarkonium production at forward rapidity in pp collisions at $\sqrt{s}=8$ TeV,''
Eur. Phys. J. C \textbf{76}, no.4, 184 (2016)
%doi:10.1140/epjc/s10052-016-3987-y
%[arXiv:1509.08258 [hep-ex]].

\bibitem{LHCb:2013itw}
R.~Aaij \textit{et al.} [LHCb],
%``Production of J/psi and Upsilon mesons in pp collisions at sqrt(s) = 8 TeV,''
JHEP \textbf{06}, 064 (2013)
%doi:10.1007/JHEP06(2013)064
%[arXiv:1304.6977 [hep-ex]].

%\bibitem{CMS:2014bsd}
%V.~Khachatryan \textit{et al.} [CMS],
%%``Measurement of the production cross section ratio $\sigma$(Xb2(1P)) / $\sigma$(Xb1(1P) in pp collisions at $\sqrt s $ = 8 TeV,''
%Phys. Lett. B \textbf{743}, 383-402 (2015)
%doi:10.1016/j.physletb.2015.02.048
%[arXiv:1409.5761 [hep-ex]].
%

\bibitem{CDF:2001fdy}
D.~Acosta \textit{et al.} [CDF],
%``$\Upsilon$ Production and Polarization in $p\bar{p}$ Collisions at $\sqrt{s}=$ 1.8-TeV,''
Phys. Rev. Lett. \textbf{88}, 161802 (2002)
%doi:10.1103/PhysRevLett.88.161802

\bibitem{Eskola:2009uj}
K.~J.~Eskola, H.~Paukkunen and C.~A.~Salgado,
%``EPS09: A New Generation of NLO and LO Nuclear Parton Distribution Functions,''
JHEP \textbf{04}, 065 (2009)
%doi:10.1088/1126-6708/2009/04/065
[arXiv:0902.4154 [hep-ph]].

\bibitem{Liu:2009wza}
Y.~Liu, Z.~Qu, N.~Xu and P.~Zhuang,
%``Rapidity Dependence of J/psi Production at RHIC and LHC,''
J. Phys. G \textbf{37}, 075110 (2010)
%doi:10.1088/0954-3899/37/7/075110
%[arXiv:0907.2723 [nucl-th]].


\bibitem{Chen:2013wmr}
B.~Chen, Y.~Liu, K.~Zhou and P.~Zhuang,
%``$\psi^\prime$ Production and $B$ Decay in Heavy Ion Collisions at {LHC},''
Phys. Lett. B \textbf{726}, 725-728 (2013)
%doi:10.1016/j.physletb.2013.09.036
%[arXiv:1306.5032 [nucl-th]].


\bibitem{Chen:2018kfo}
B.~Chen,
%``Thermal production of charmonia in Pb-Pb collisions at $\sqrt {s_{NN}} = $ 5.02 TeV,''
Chin. Phys. C \textbf{43}, no.12, 124101 (2019)
%doi:10.1088/1674-1137/43/12/124101
%[arXiv:1811.11393 [nucl-th]].

\bibitem{Shi:2017qep}
W.~Shi, W.~Zha and B.~Chen,
%``Charmonium Coherent Photoproduction and Hadroproduction with Effects of Quark Gluon Plasma,''
Phys. Lett. B \textbf{777}, 399-405 (2018)
%doi:10.1016/j.physletb.2017.12.055
%[arXiv:1710.00332 [nucl-th]].

\bibitem{ALICE:2018wzm}
S.~Acharya \textit{et al.} [ALICE],
%``$\Upsilon$ suppression at forward rapidity in Pb-Pb collisions at $\sqrt{s_{\rm NN}}$ = 5.02 TeV,''
Phys. Lett. B \textbf{790}, 89-101 (2019)
%doi:10.1016/j.physletb.2018.11.067
%[arXiv:1805.04387 [nucl-ex]].


\bibitem{CMS:2017ycw}
A.~M.~Sirunyan \textit{et al.} [CMS],
%``Suppression of Excited $\Upsilon$ States Relative to the Ground State in Pb-Pb Collisions at $\sqrt{s_\mathrm{NN}}$=5.02 TeV,''
Phys. Rev. Lett. \textbf{120}, no.14, 142301 (2018)
%doi:10.1103/PhysRevLett.120.142301
%[arXiv:1706.05984 [hep-ex]].


\bibitem{Du:2017qkv}
X.~Du, R.~Rapp and M.~He,
%``Color Screening and Regeneration of Bottomonia in High-Energy Heavy-Ion Collisions,''
Phys. Rev. C \textbf{96}, no.5, 054901 (2017)
%doi:10.1103/PhysRevC.96.054901
%[arXiv:1706.08670 [hep-ph]].


\bibitem{Ye:2017fwv}
Z.~Ye [STAR],
%``$\Upsilon$ measurements in p+p, p+Au and Au+Au collisions at $\sqrt {s_{NN}}$ = 200GeV with the STAR experiment,''
Nucl. Phys. A \textbf{967}, 600-603 (2017)
%doi:10.1016/j.nuclphysa.2017.06.040

\bibitem{STAR:2022rpk}
 [STAR],
%``Observation of sequential $\Upsilon$ suppression in Au+Au collisions at $\sqrt{s_{_\mathrm{NN}}}$ = 200 GeV with the STAR experiment,''
[arXiv:2207.06568 [nucl-ex]].

\end{thebibliography}
\end{document}